\documentclass[times,11pt,psfig,twoside]{actaprep}
\usepackage{graphicx}
\usepackage{pslatex}
\usepackage{subfig}
\usepackage{longtable}
\usepackage{caption}
\SetPages{001}{020}
\SetVol{66}{2016}

\begin{document}

\captionsetup{font=small}
\tabcolsep 4pt

\begin{Titlepage}
\Title{OGLE Collection of Star Clusters. \\
New Objects in the Outskirts of the Large Magellanic Cloud}

\Author{~~~M.~~S~i~t~e~k$^1$, ~~~M.~K.~~S~z~y~m~a~{\'n}~s~k~i$^1$,
  ~~~D.~M.~~S~k~o~w~r~o~n$^1$, ~~~A.~~U~d~a~l~s~k~i$^1$,
  ~~~Z.~~K~o~s~t~r~z~e~w~a~-~R~u~t~k~o~w~s~k~a$^{2,3}$, ~~~J.~~S~k~o~w~r~o~n$^1$,
  ~~~P.~~K~a~r~c~z~m~a~r~e~k$^1$, ~~M.~~C~i~e~{\'s}~l~a~r$^1$,
  ~~~{\L}.~~W~y~r~z~y~k~o~w~s~k~i$^1$, ~~~S.~~K~o~z~{\l}~o~w~s~k~i$^1$,
  ~~~P.~~P~i~e~t~r~u~k~o~w~i~c~z$^1$, ~~~I.~~S~o~s~z~y~{\'n}~s~k~i$^1$,
  ~~~P.~~M~r~{\'o}~z$^1$, ~~~M.~~P~a~w~l~a~k$^1$, ~~~R.~~P~o~l~e~s~k~i$^{1,4}$,
  ~~~K.~~U~l~a~c~z~y~k$^{1,5}$}
  {$^1$Warsaw University Observatory, Al~Ujazdowskie~4, 00-478~Warszawa, Poland\\
   $^2$SRON Netherlands Institute for Space Research, Sorbonnelaan 2, 3584 CA Utrecht, the Netherlands\\
   $^3$Department of Astrophysics/IMAPP, Radboud University Nijmegen, P.O. Box 9010, 6500 GL Nijmegen, the Netherlands\\
   $^4$Department of Astronomy, Ohio State University, 140 W. 18th Ave., Columbus, OH 43210, USA\\
   $^5$Department of Physics, University of Warwick, Gibbet Hill Road, Coventry, CV4 7AL, UK\\
e-mail:(msitek, msz)@astrouw.edu.pl}

\Received{~~}
\end{Titlepage}

\Abstract{The Magellanic System (MS), consisting of the Large Magellanic Cloud (LMC),
the Small Magellanic Cloud (SMC) and the Magellanic Bridge (MBR), contains
diverse sample of
  star clusters. Their spatial distribution, ages and chemical
  abundances may provide important information about the history of formation 
of the whole System. 
We  use deep photometric maps derived from the images collected during the fourth phase of The
  Optical Gravitational Lensing Experiment (OGLE-IV) to construct the
  most complete catalog of star clusters in the Large Magellanic Cloud
  using the homogeneous photometric data. In this paper we
  present the collection of star clusters found in the area of about
  225 square degrees in the outer regions of the LMC. Our sample
  contains 679 visually identified star cluster candidates, 226 of
  which were not listed in any of the previously published catalogs.
  The new clusters are mainly young small open clusters or clusters
  similar to associations. 
}{Catalogs Star Clusters: general Surveys}

\section{Introduction}

The Magellanic System is an ideal astrophysical laboratory for studying
the structure and evolution of galaxies. It is located close enough to the
Galaxy so that millions of stars can be easily resolved.
This can be used in a range of scientific projects exploring 
stellar populations, their evolution, ages, metallicity
(Jacyszyn-Dobrzeniecka \etal 2016, Skowron \etal 2014). 
One of the methods for such studies is the investigation of star clusters. The
Large Magellanic Cloud contains a large sample of these
systems. The spatial distribution od clusters, their age, chemical composition,
structural parameters and dynamical evolution may
provide valuable information about the LMC formation history.
The Magellanic System has a very rich and diverse structure. The
positions of centroids of both Clouds depend on which stellar
population is used for their estimation (Cioni, Habing, Israel 2000, 
Deb and Singh 2014). Moreover, the
Magellanic Clouds 
look asymmetrical, with denser parts located
toward the Bridge (Klein \etal 2014, Scowcroft \etal 2015). Analysis of the
spatial distribution of the star clusters in the MS enables the comparison
to the general stellar populations. This may give
some hints of what caused the asymmetry of the LMC and the SMC. The correlation between
age, size, metallicity and spatial distribution can bring new
information about the MS history (Palma \etal 2016, Piatti
\etal 2014). To make all these studies possible, however, a complete
collection of star clusters is needed, derived from homogeneous
observational data, possibly from a single photometric survey.

So far, the largest catalog of extended objects
(excluding background galaxies) in the Magellanic System was published
by Bica \etal (2008) and
contains a compilation of all the
previously published catalogs. The most important contribution to this sample was due to
the catalog based on the OGLE-II data published by Pietrzy{\'n}ski \etal
(1999). That catalog, however, covered only the central
part of the LMC: 5.8 square degrees -- only 3\% of the area
observed during the current OGLE-IV phase (Udalski, Szyma{\'n}ski and
Szyma{\'n}ski 2015).

The parameters (such as size, age, metalicity) of some clusters 
listed in the Bica catalog (Bica \etal 2008) were
estimated by several groups, including Glatt, Grebel and Koch (2010),
Palma \etal\ (2016), Choudhury, Subramaniam, Piatti (2015) and Piatti
\etal\ (2002, 2003a,b, 2009, 2014, 2015). The OGLE survey data were
also used for this purpose, first by Pietrzy{\'n}ski and Udalski (2000)
using OGLE-II data (Udalski, Kubiak and Szyma{\'n}ski, 1997) and recently,
by Nayak \etal\ (2016), using OGLE-III data (Udalski \etal 2008).

The OGLE-IV survey covers practically the entire
Magellanic System inclu\-ding both galaxies and the Magellanic Bridge
(about 650 square degrees). The large collection of high resolution
individual images enabled constructing deep images of the
entire region with a depth comparable to images collected by a 4-meter
class telescope. Uniform OGLE-IV photometric data provides a unique,
homogeneous data set for the search of star clusters and
associations. It will allow us to create a complete catalog of star
clusters in the MS and their basic parameters.

This paper presents the first part of the catalog based on
the OGLE-IV data. The central part of the LMC has already been
observed or analyzed by many other projects so we decided to start our
exploration with the outer parts of the galaxy.

We have found 679 star clusters in the outer regions of the LMC.
Among those, 226 objects were not listed in any of the previous catalogs,
438 objects were listed in the Bica catalog and 15 objects were listed in the Dark
Energy Survey (DES) publication (Pieres \etal 2016).
As some extended objects cannot be unambiguously classified, we have
 performed a cross-match of our sample to both star clusters
and associations from the Bica catalog. 
Almost all known objects which were in the area of
the analyzed OGLE fields were detected by our algorithm, proving the
effectiveness of algorithm and the completeness of the sample. There are five Bica
objects which we do not see in our images and one DES object which 
is located in the gap between the OGLE subfields.

\section{Observations and Data Reduction}

The photometric data of the LMC fields analyzed in this paper are based
on the images gathered during the first five years (2010--2015) of the
fourth phase of the OGLE project (OGLE-IV, Udalski, Szyma{\'n}ski, and
Szyma{\'n}ski 2015). The project has been conducted on the 1.3-meter Warsaw
Telescope at the Las Campanas Observatory, equipped with the 32-chip
mosaic CCD camera, covering 1.4 square degree field with 0.26 arcsec per
pixel scale (see the aforementioned paper for the details of the
OGLE-IV hardware setup, observational strategy and photometric
reductions pipeline).

We have used the ``deep photometric maps'' -- catalogs of all objects
detected on the deep images of all observed fields. 
The deep photometric maps for the OGLE-IV
have not been published yet, but they were prepared in a similar way to
those published for the OGLE-III phase of the project (see Udalski
\etal 2008 for the LMC OGLE-III data). The main, crucial
difference is that the OGLE-IV maps are much deeper, as the template
images were constructed, for the $I$-band, by stacking up to 100
images taken in the best photometric weather conditions. For all 213
observed LMC fields, the number of the stacked images used for the templates is
between 61 and 100 (90 on average), depending on the overall number of
good seeing individual
images available for any given field. For $V$-band, which is observed
less frequently, the templates were constructed from 4 to 100
individual images with the mean value of 19. For comparison, the
templates for 116 OGLE-III LMC fields were constructed of at most 10 images.
We refer to the OGLE-IV maps used in this paper as ``deep photometric
maps'' in order to distinguish them from the preliminary maps
constructed for internal use in 2012 from much ``shallower''
templates.
The catalog data includes the coordinates of objects, calibrated $V$ and $I$ mean magnitudes,
their errors, $V-I$ color index. 
\begin{figure}
\centering
\subfloat[]{\label{a}
\includegraphics[angle=270,width=0.45\textwidth]{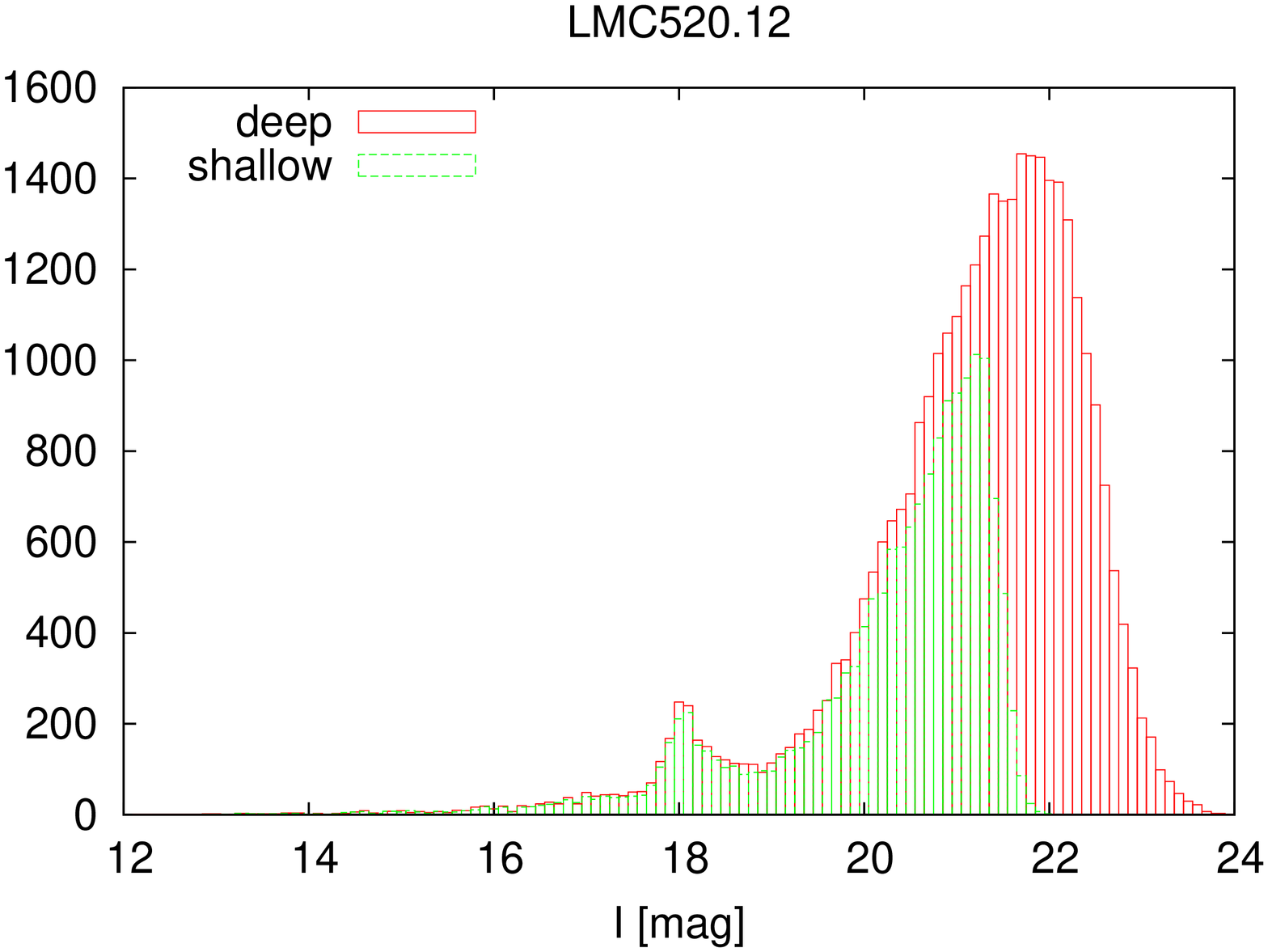}}
\quad
\subfloat[]{\label{b}
\includegraphics[angle=270,width=0.45\textwidth]{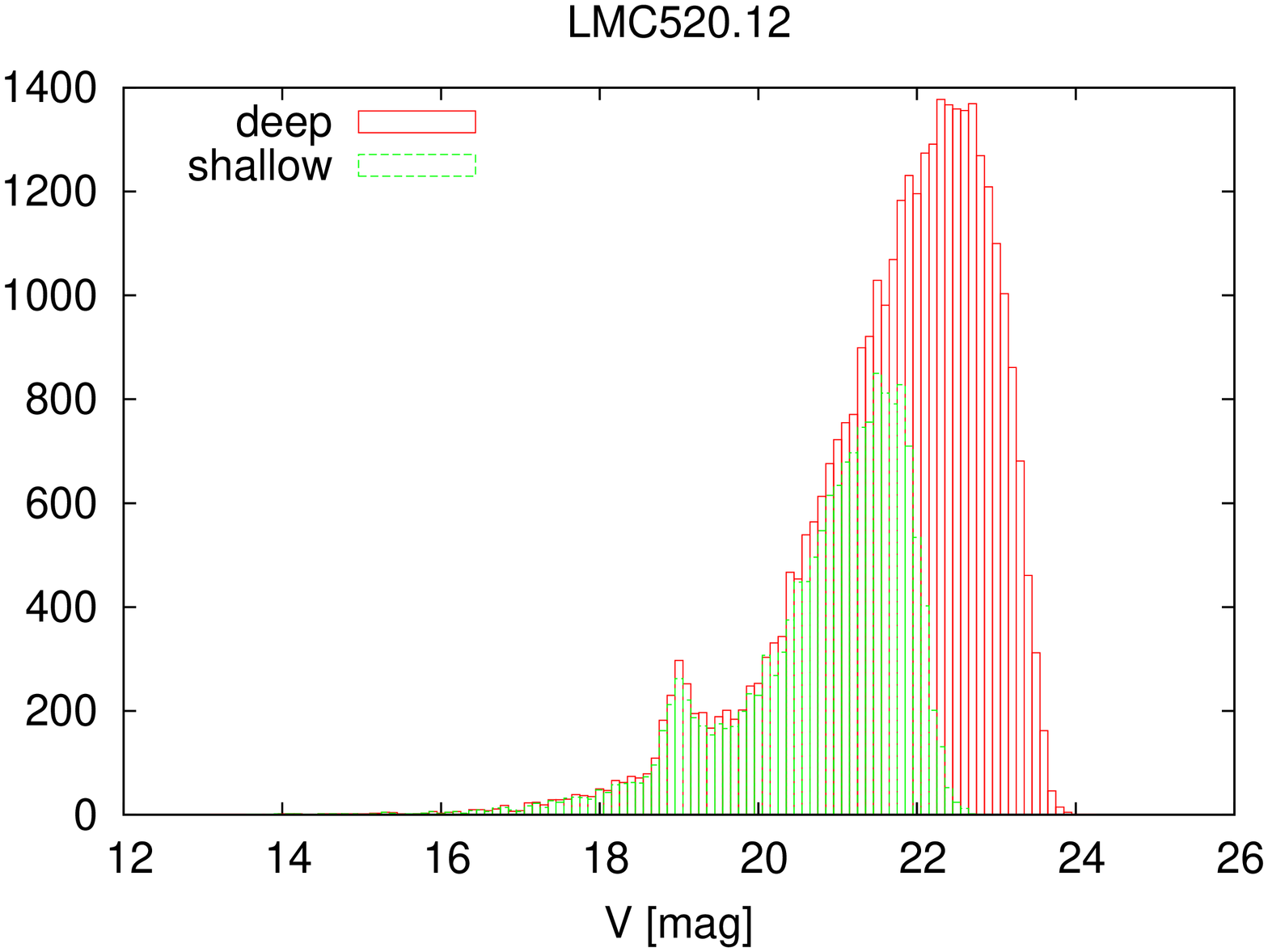}}
\quad
\subfloat[]{\label{c}
\includegraphics[angle=270,width=0.45\textwidth]{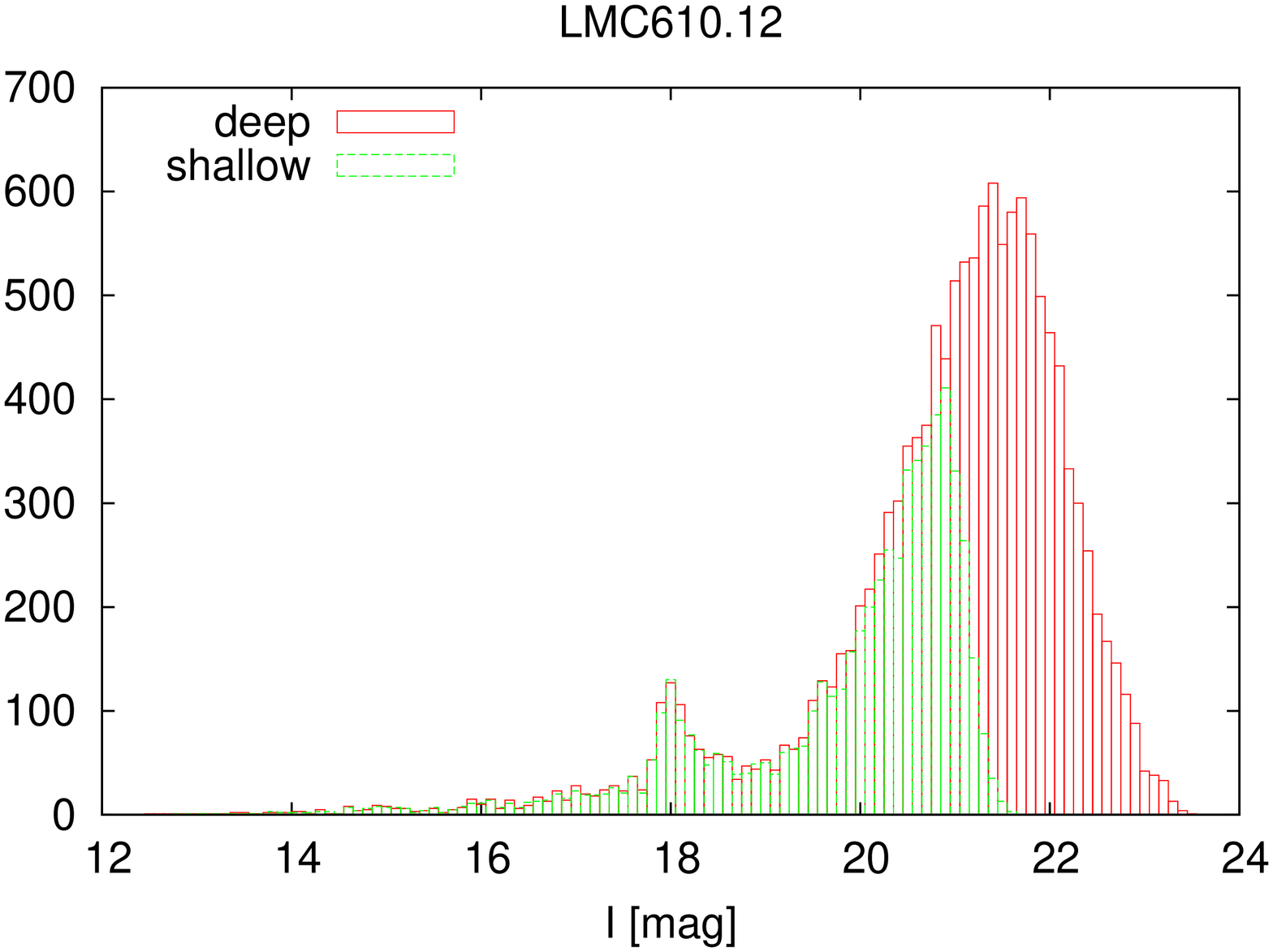}}
\quad
\subfloat[]{\label{d}
\includegraphics[angle=270,width=0.45\textwidth]{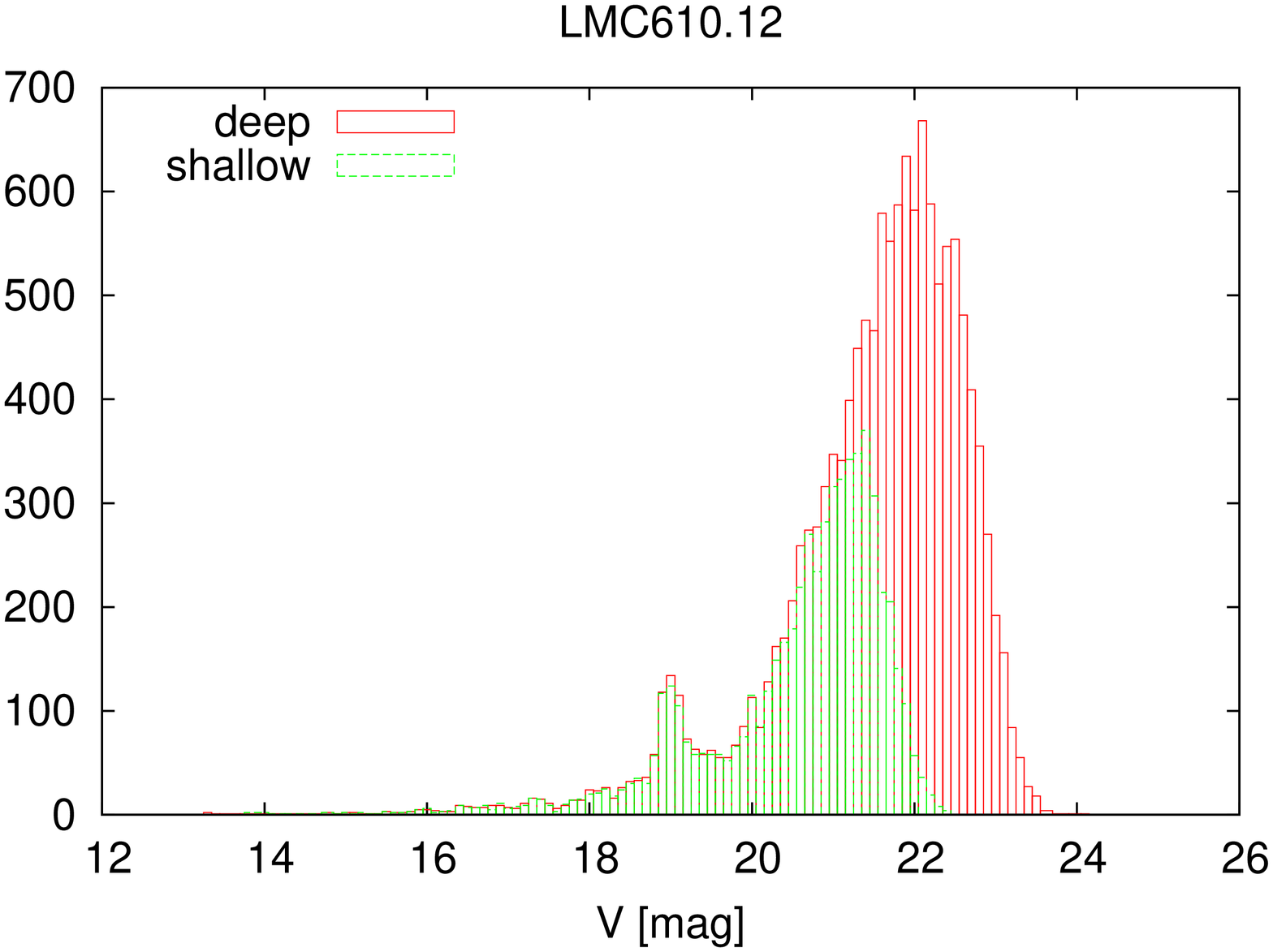}}
\caption{Histograms of brightness (left panels are $I-$band and right panels are $V-$band) for two OGLE-IV subfields LMC520.12 (a and b) and LMC610.12 (c and d).
\label{fig1}}
\end{figure}

Every observed field is divided into 32 subfields according to the
setup of the mosaic CCD camera chips. Photometry reductions, databases and
photometric maps are done independently for each subfield.

The star detection limit of the deep photometric maps of the LMC
reaches $I\approx 23.5$~mag and $V\approx 24$~mag.
The maps are complete to about 20.5--21.5 mag
in the $I$-band and 21.5--22.5 mag in the $V$-band, depending on the
location and crowding of the field. These limits are determined from 
the histograms of the star magnitudes by the estimation of the magnitude
where the numbers start to deviate from the systematic growth
(Fig. \ref{fig1}). This photometric range should allow us to detect 
a vast majority of
star clusters in the Magellanic System, with a possible exception of
the most sparse objects and objects which are located in the gaps between the fields.
To obtain such a deep range, the PSF photometry
was done directly on the template images, using  DoPHOT program
(Schechter, Mateo and Saha 1993). This is another
important difference to the previously published OGLE
photometric maps which were based on the OGLE standard pipeline
photometry obtained by using the image difference technique.

\section{Search for Clusters}

The first attempt to search for star clusters using a fully automated
algorithms was undertaken by Bhatia and MacGillivray (1989).
They found 284 clusters in a 6 square degrees region toward 
the North-East end part of the LMC bar, four times more 
than it was previously known in the observed region.
Zaritsky \etal (1997) searched for star clusters similarly as Bhatia 
and MacGillivray but they were using, for the first time, observations
from a CCD camera. The result increased the number of identified clusters 
by about 45\% in the central $8^{\circ} \times 8^{\circ}$ region of the LMC.

In general, Zaritsky's automated method is based on photometric maps which
are divided into small cells. The size of the cell depends on 
the region crowding -- the denser field is the smaller cells are needed. 
Afterwards the stars are counted in each cell and
then the algorithm looks for cells which have more stars than a
typical cell (threshold). Zaritsky's method is still in use with small
modifications depending on the quality of observations. For example
Pietrzy{\'n}ski \etal (1999) used this method as well on the OGLE-II
data.

Our search for star clusters was based on the deep photometric maps
filtered effectively from most artefacts by using only objects
detected in both I- and V-band deep template images. We have divided the LMC
region into two parts -- the disk region and the outer part of the LMC.
Here, we present the analysis of the fields located outside the LMC
disk (Fig.\ 2). Our method is effective and reliable only in sparse
stellar fields. The search method for the densest fields will be
described in the next part of the catalog.

\begin{figure}
\includegraphics[width=1\textwidth]{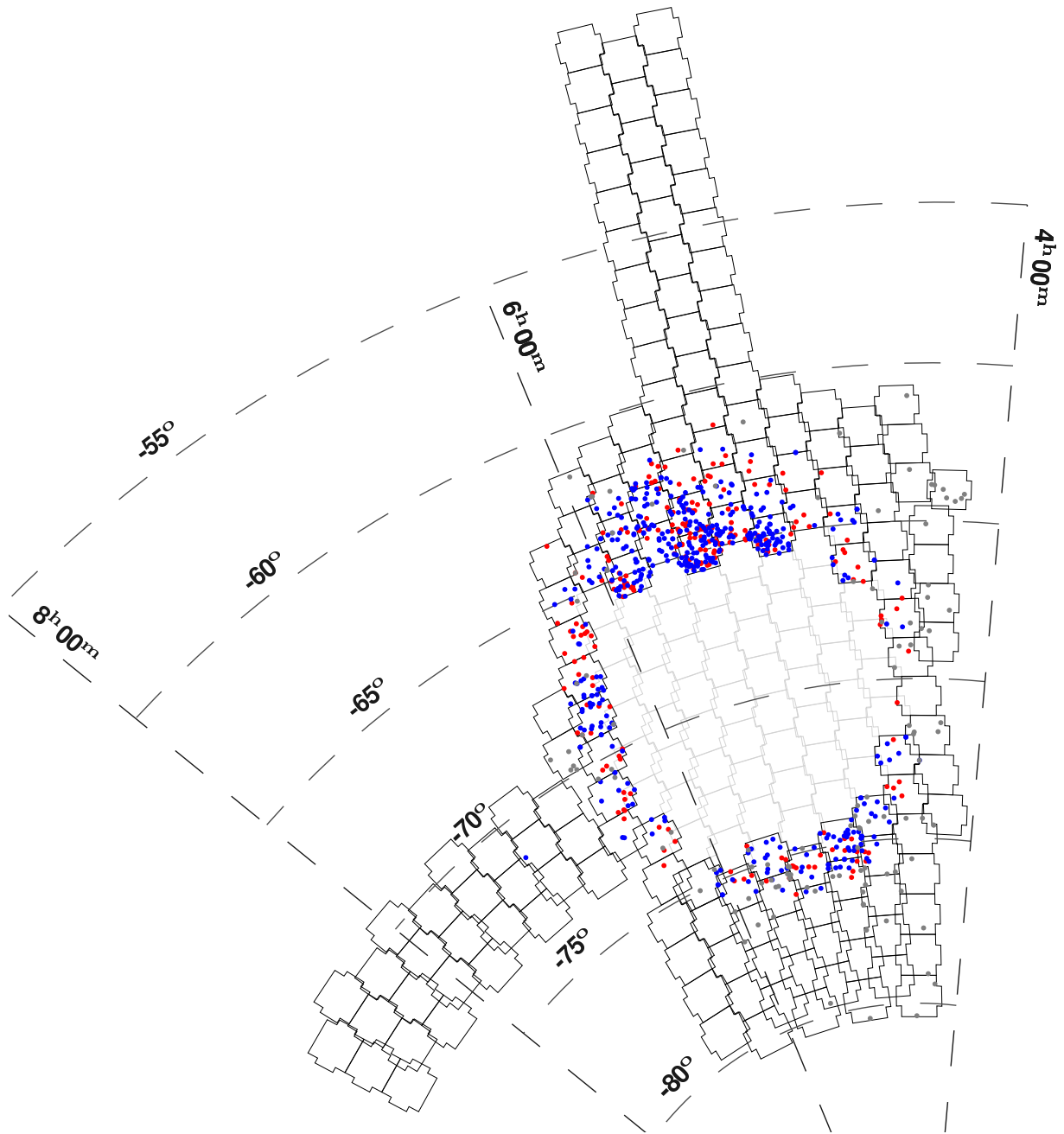}
\caption{The OGLE-IV fields in the LMC region. Outer black polygons
  were analyzed in this paper. Red and blue
  dots mark the newly discovered and the previously known star
  clusters, respectively. Gray dots are objects which passed the visual 
inspection but were rejected by the reliability cut off (see Section 4).}
\label{fig2-fields}
\end{figure}

The examined area of 225 square degrees contains 165 OGLE-IV fields
(5280 single subfields). 
The analyzed region includes six denser fields
(LMC513, LMC520, LMC564, LMC73, LMC592, LMC596) located
north of declination $-65^{\circ}$, to allow full comparison with 
the DES results.
All analyzed fields are shown in Fig. \ref{fig2-fields}
marked with black polygons. The gray polygons mark the central LMC
fields which have not been analyzed here. The list of all analyzed LMC
fields and their central coordinates is available on the Web page
together with other supplementary information (see Section 4 for details).

\begin{figure}
\centering
\subfloat[counts in 47 pixels (12") cells]{\label{oldA}
\includegraphics[angle=270,width=0.45\textwidth]{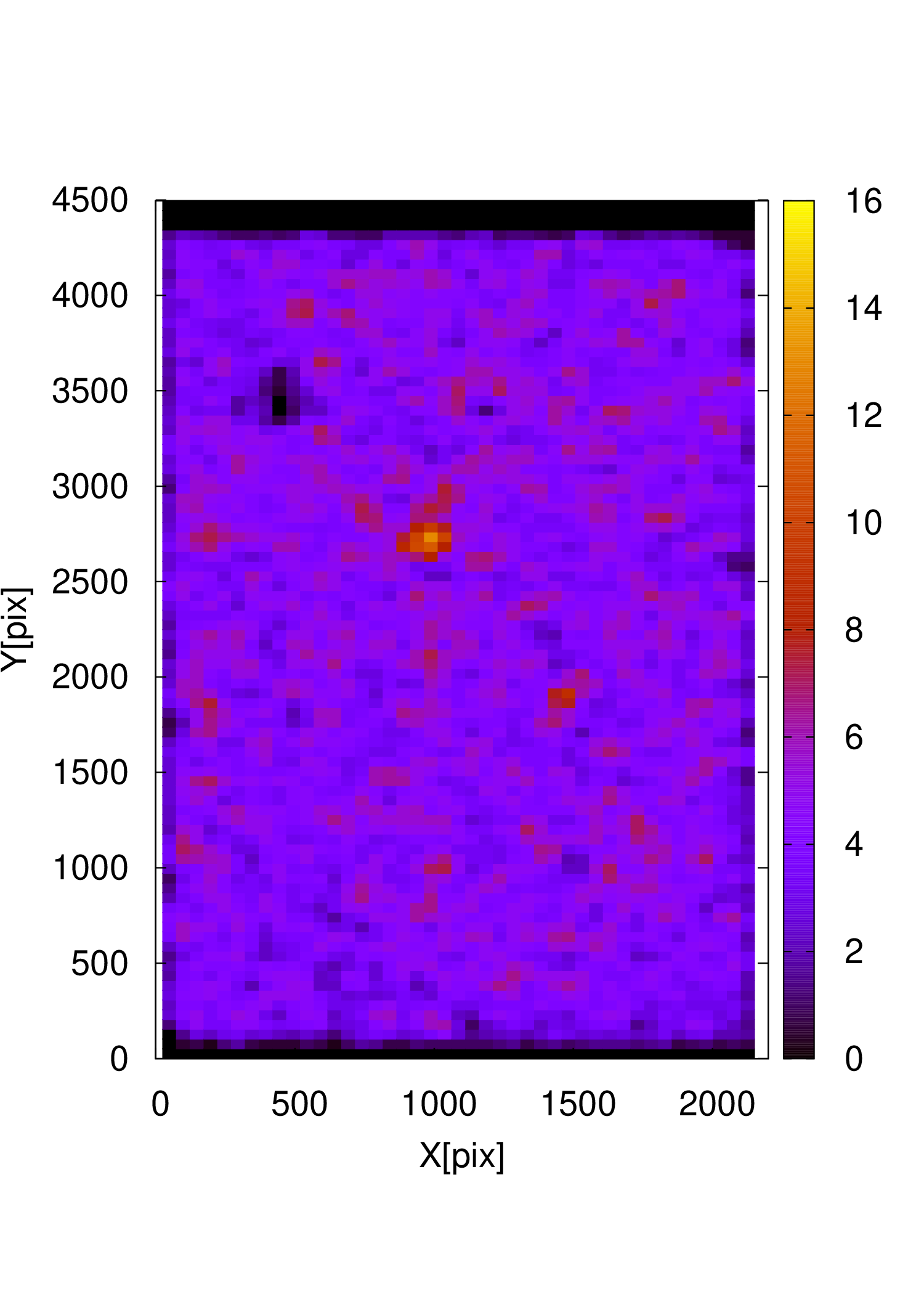}}
\quad
\subfloat[counts in 23 pixels (6") cells]{\label{oldB}
\includegraphics[angle=270,width=0.45\textwidth]{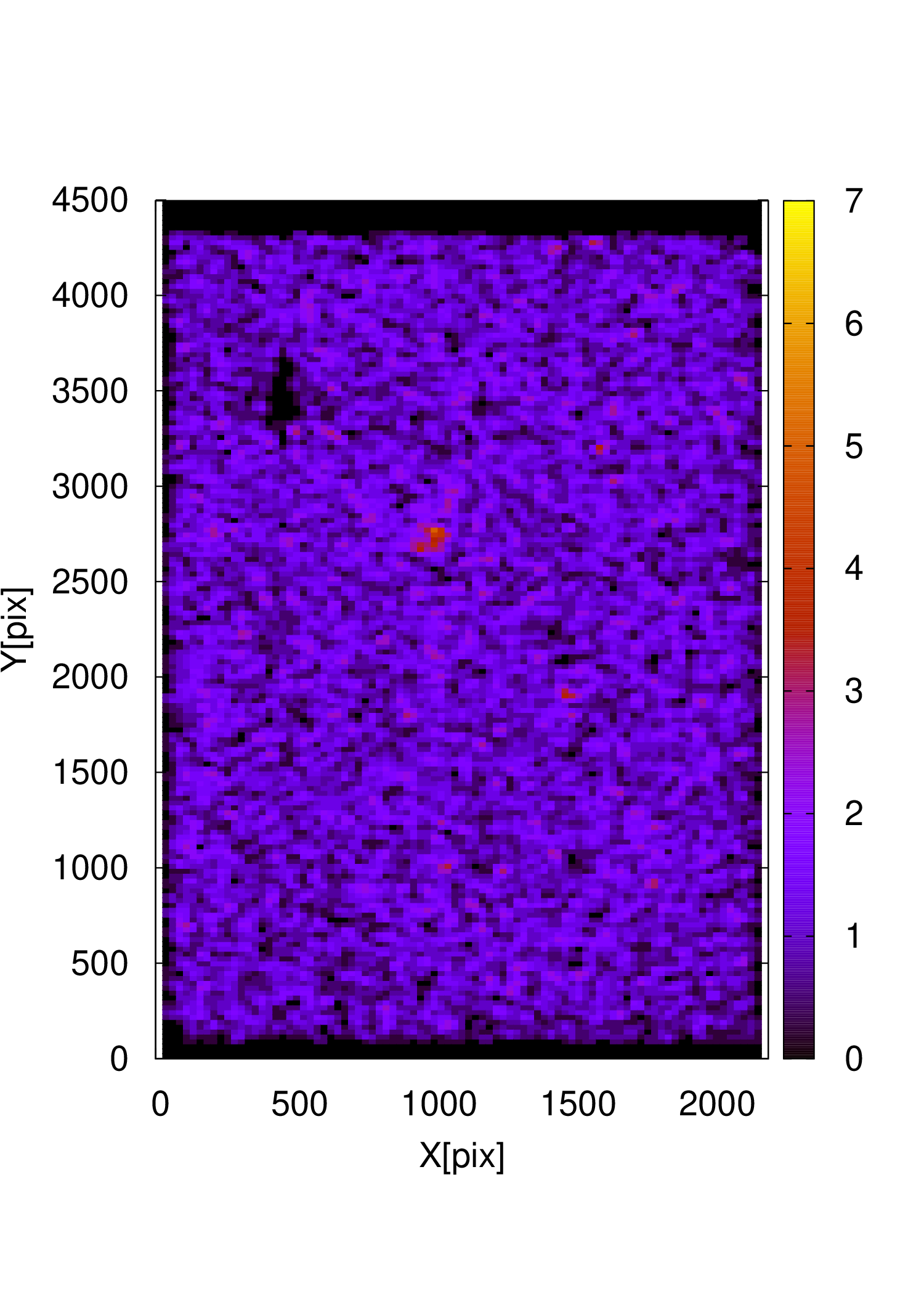}}
\caption{ Stellar density maps of the LMC610.07 field for two
  different cell sizes. The field contains two clusters: OGLE-LMC-CL-0883, 
which is 
located around (1490,1900) and, previously known, OGLE-LMC-CL-1025
(SL873, LW447, KMHK1715) around (1000,2750).
\label{fig3}}
\end{figure}

\begin{figure}
\centering
\subfloat[counts in 47 pixels (12") cells]{\label{zuziaA}
\includegraphics[angle=270,width=0.45\textwidth]{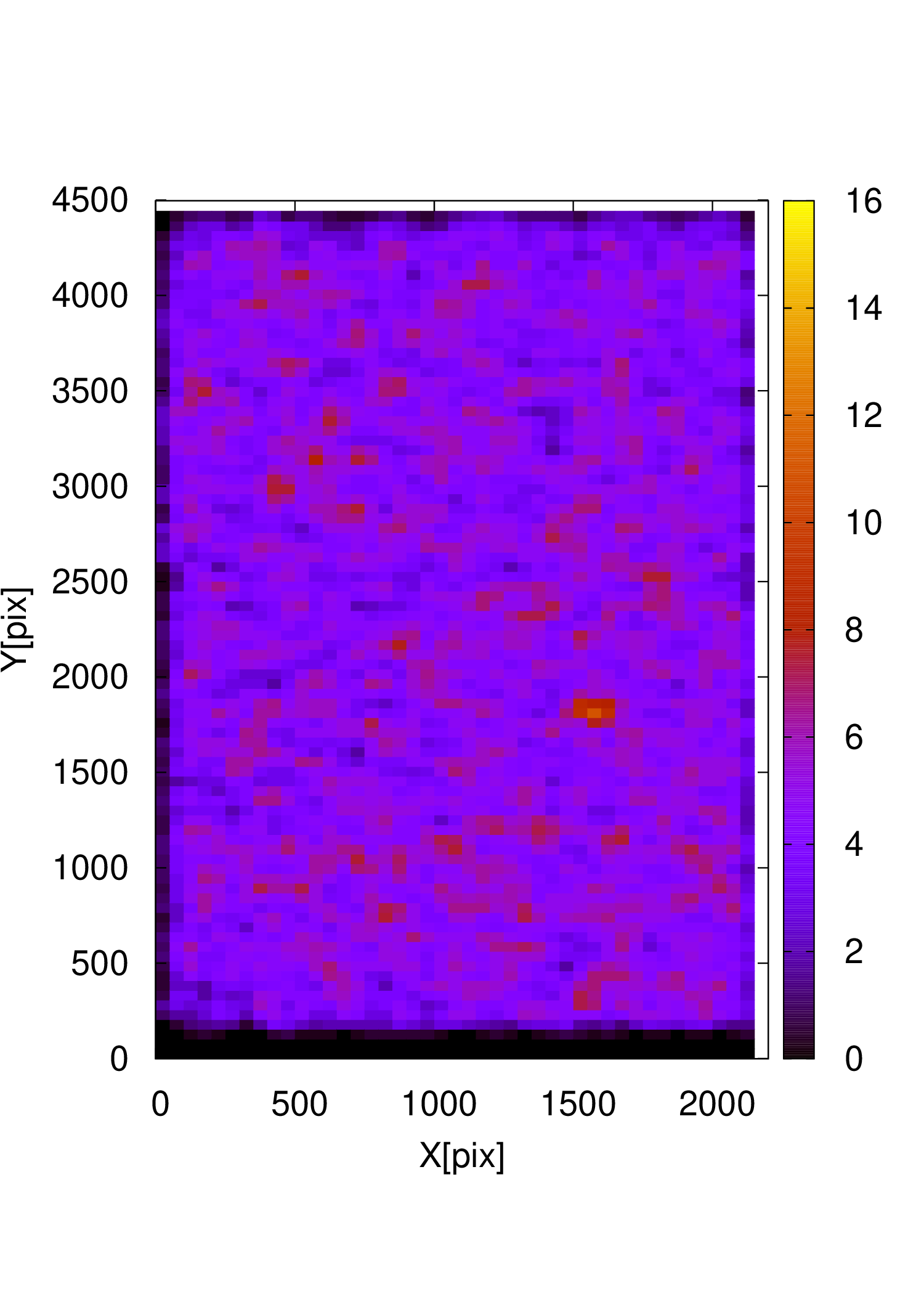}}
\quad
\subfloat[counts in  23 pixels (6") cells]{\label{zuziaB}
\includegraphics[angle=270,width=0.45\textwidth]{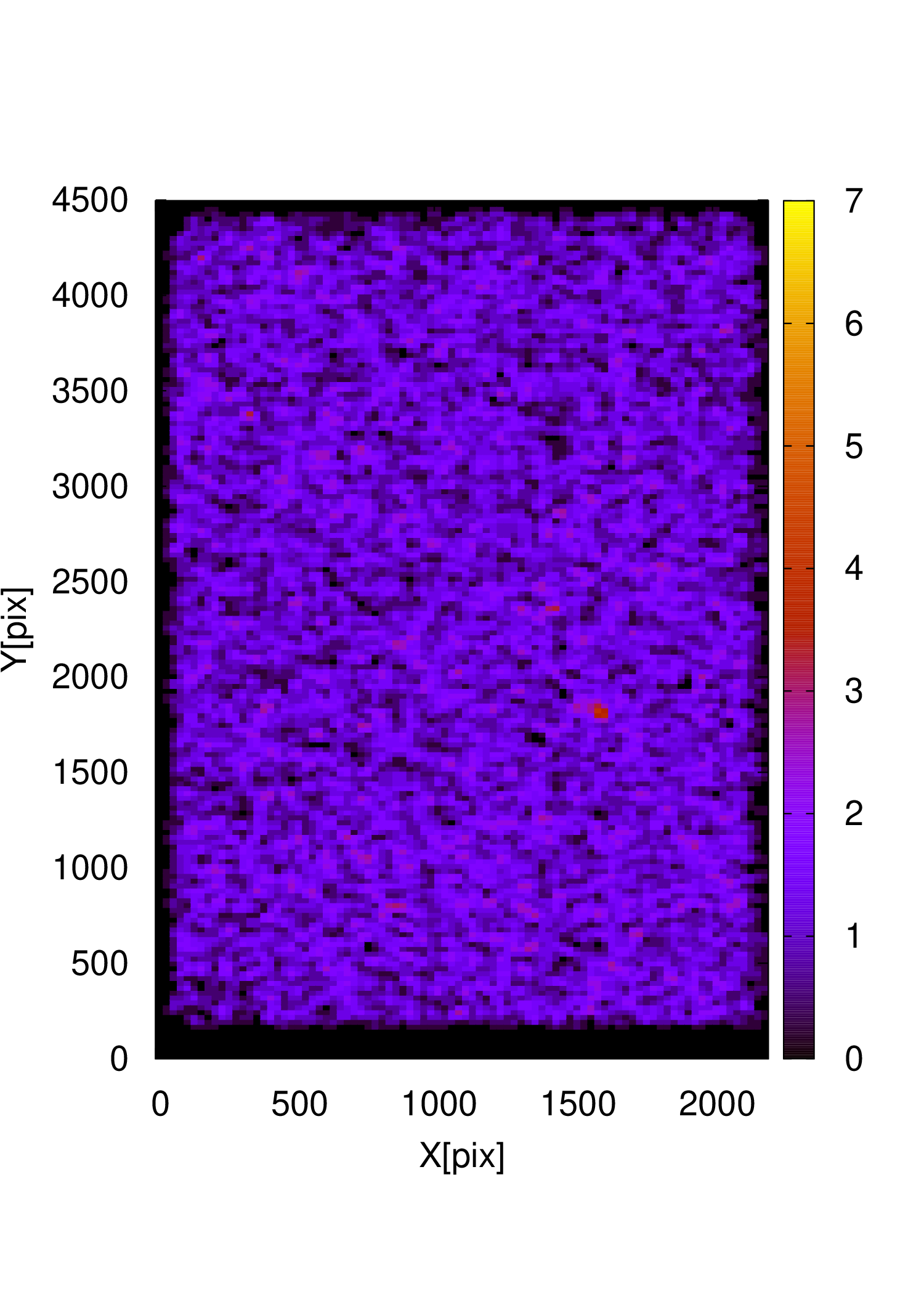}}
\caption{The stellar density maps of a subfield number LMC582.01 with object OGLE-LMC-CL-0877 located around (1600,1850).
\label{fig4}}
\end{figure}

We applied the search method as follows.
First, we divided each photometric map into square cells and counted
the objects in each cell to create a density map. For each field we
prepared three density maps with different cell sizes -- 23~pixels
(6"), 47~pixels (12") and 93~pixels (24"). Maps with the smallest cells were
suitable for finding the smallest clusters or clusters similar
to associations, while the bigger cells were suitable only for globular
clusters or large and dense open clusters. Exemplary density maps are
presented in Figs. \ref{fig3} and \ref{fig4}.

Outer regions of the LMC are generally sparse.
Still, the stellar density varies significantly, so we
divided the density maps into three groups with a different threshold
values. The threshold is a value of counts in a cell above which our 
algorithm selected the densier areas in each subfield. 
The thresholds were based on median counts in cells with the size of
47~pixels. The first group contained the most dense fields where the
value of the median was larger than a half of the value of the maximum
count in the field. In this group, we chose the cells with the counts
higher than the median values plus one. In the second group we defined
the threshold as two times the median value plus two. The third group contains
the most sparse fields. In those we selected the cells with counts
higher than the half of the maximum value. Those three density groups
and their threshold values are defined in Table
\ref{tabela1-beseline}.

\begin{table}
\begin{center}
\caption{Cell density groups and their threshold values. The \textit{median}
  and \textit{maximum} denote the number of counts in a cell}
\begin{tabular}{|c|c|c|c|} \hline
group & I ($median>\frac{maximum}{2}$) & II ($0<median<\frac{maximum}{2}$)  & III ($median=0$) \\ \hline
threshold & $median+1$ & $2*median + 2$ & $\frac{maximum}{2}$ \\ \hline
\end{tabular}
\label{tabela1-beseline}
\end{center}
\end{table}

The cells with values higher than the threshold which were next to each
other were merged to one -- the cell
with the highest count. The coordinates
of this cell were used as the first estimation of the central point of
the star concentration in the area.

\begin{figure}
\includegraphics[width=1\textwidth]{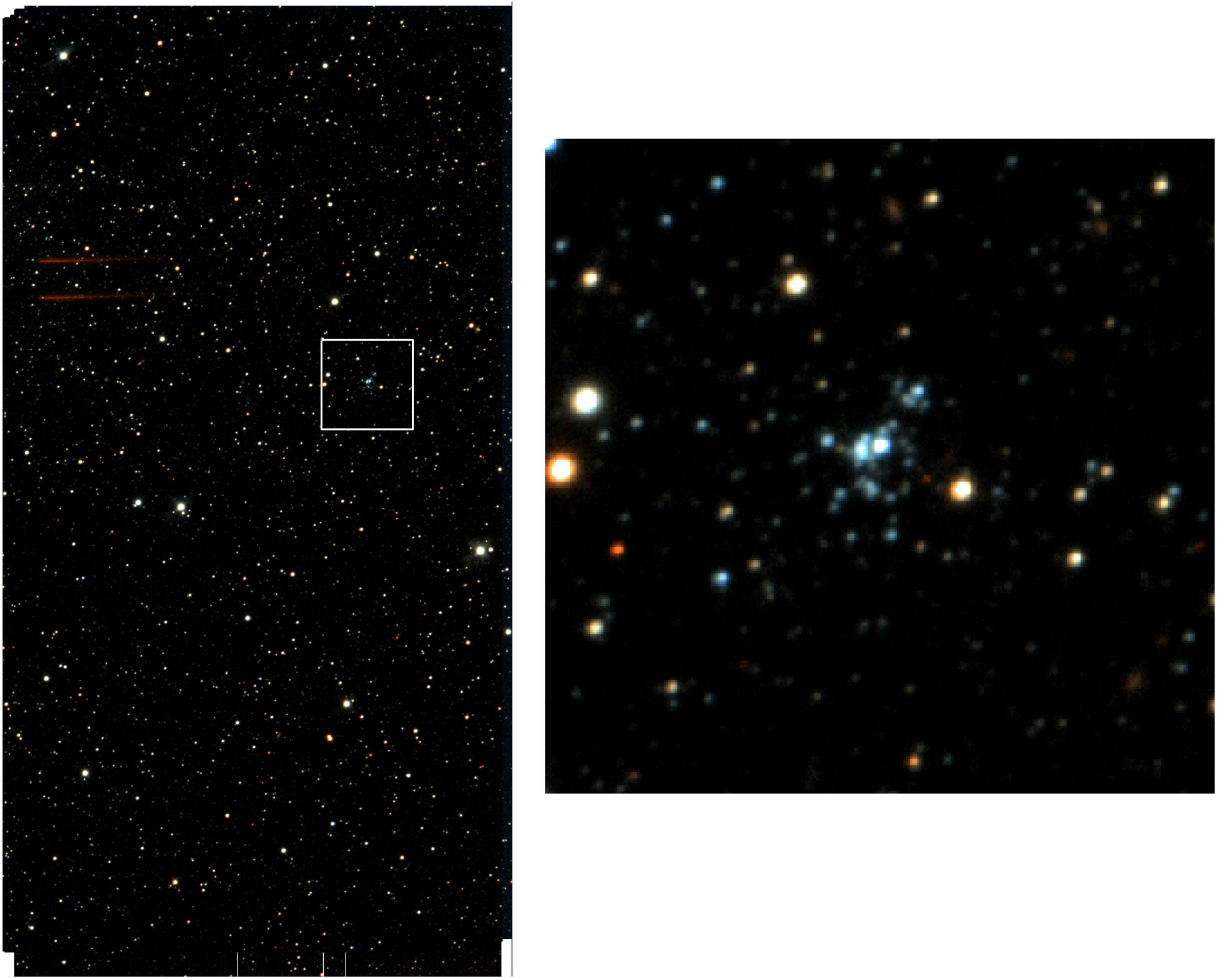}
\caption{Color image of the field LMC582.01, The white square
  ($1.7'\times 1.7'$) is enlarged in the right panel, clearly
  showing the cluster OGLE-LMC-CL-0877.}
\label{fig5-kolor}
\end{figure}

\begin{figure}
\centering
\subfloat[]{\label{fig6a-fotometric}
\includegraphics[width=0.45\textwidth]{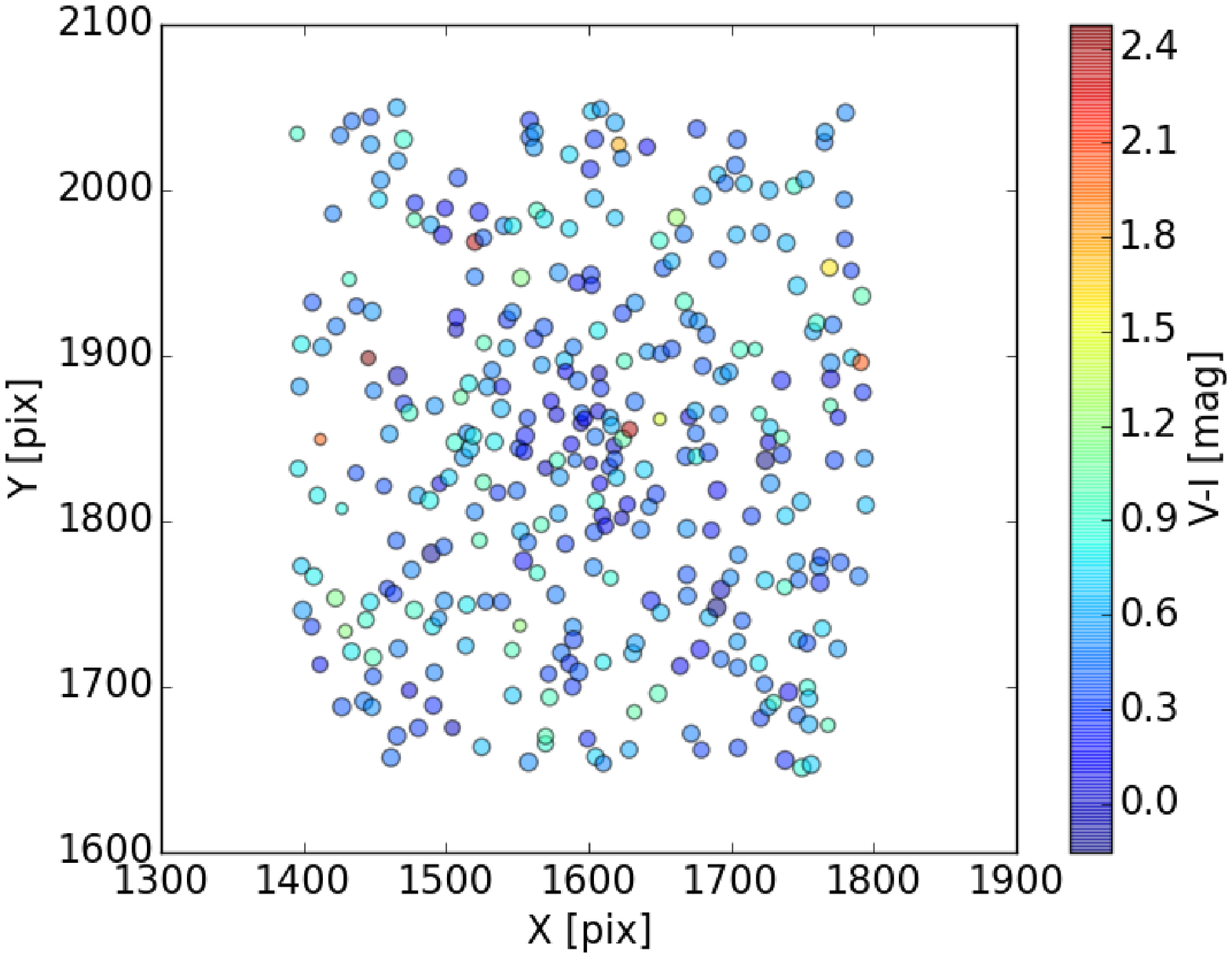}}
\quad
\subfloat[]{\label{fig6b-kde}
\includegraphics[width=0.45\textwidth]{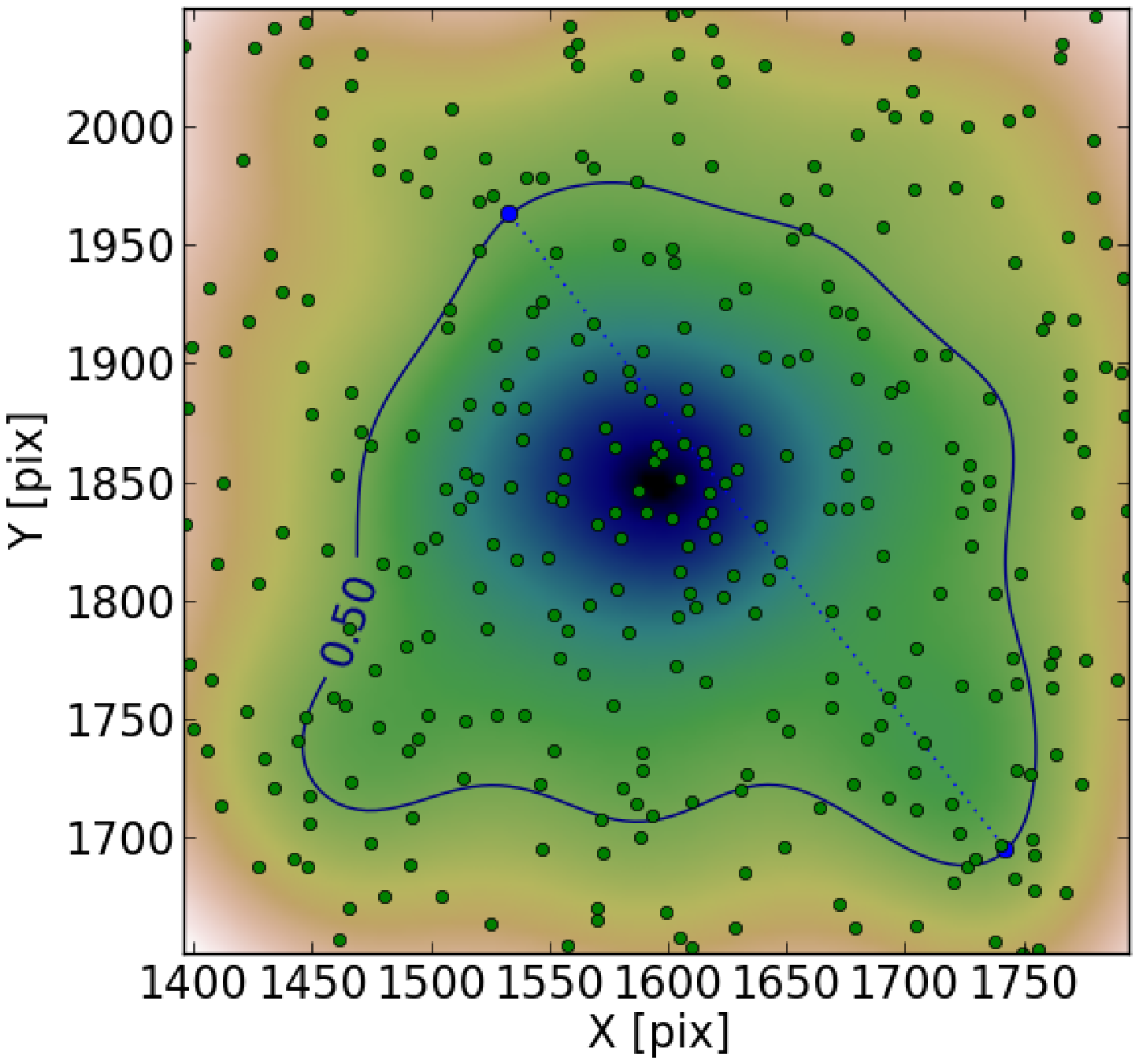}}
\caption{Deep photometric map of the cluster
  OGLE-LMC-CL-0877 with the same size as in Fig. \ref{fig5-kolor} --
  $1.7'\times 1.7'$.
  (a) The size of each point is proportional to the
  brightness of a star in $I$-band, and the color represents
  the $V-I$ color of each star. (b) Deep photometric map with
  standard KDE distribution which was used to estimate
  the object's centroid. }
\label{fig6}
\end{figure}

The algorithm yielded amost 6000 candidate areas which were visually inspected by 
cutting out regions of 400$\times$ 400 pixels (1.7 $\times$ 1.7 arcmin). For every region we
constructed a false-color composition of images taken in $I-$ and
$V-$ band (Fig. \ref{fig5-kolor}) and plotted a photometric map with
marked brightness and color ($V-I$) of the stars (see Fig.
\ref{fig6a-fotometric} for a sample plot).

Most of the false detections were caused by bright and overexposed
stars, background galaxies spuriously detected by the PSF software as
groups of stars, and other outliers on the edge of the field. It is
worth noting that due to the way of the construction of the template
images, the quality of the photometry of objects close to the edges
of the subfields is worse than in other regions, due to the smaller
number of individual components in the template image.

The visual inspection was based on the density maps, color images,
template images in both filters and photometric maps. We compared all
seven pictures (three density maps, $I$-band image, $V$-band image,
composition of $I-$ and $V-$ band images, photometric map) for each
potential star cluster. If the cluster was visible on all of them,
the reliability index of detection was set to one, but for each picture
not showing any concentration of stars this parameter
was reduced by 0.1. If the assigned reliability value was smaller
than 0.6, the object was rejected. 
Of all objects which passed the visual inspection and were
assigned the reliability index, 59\% received the maximum value of 1, 5\% -- 0.9,
3\% -- 0.8, 4\% -- 0.7 and 6\% -- 0.6. The remaining 23\% with 
the index less than 0.6
were rejected from the sample.
All pictures (images in $V-$ and
$I-$ band, color images and a photometric map) of the accepted star
clusters are shown on the Web page (Section 4).

To determine the centroids of the star clusters we used standard
Gaussian Kernel Density Estimation (KDE) from the Python scipy.stats
library. Firstly, we selected the stars from a 400 $\times$
400 pixels square around the central point estimated from the densest cell.
Then we computed the coordinates of the maximum value from KDE and
again we selected the stars from the square of the same size centered
on the point of the KDE maximum value. We repeated the procedure three
times to obtain acceptable accuracy of the centroid. The subsequently 
computed coordinates differ by not more than a
few pixels ($\sim 1$ arcsec). Exemplary KDE for the object OGLE-LMC-CL-0877 is shown
in Fig. \ref{fig6b-kde}.

To estimate the approximate size of a star cluster we 
calculated the average distance from the centroid to the KDE contour line
located at half maximum value (normalized to 0.5, see Fig. \ref{fig6b-kde} 
for a sample contour).

All the centroids were calculated in XY coordinates of the field and
then converted to equatorial coordinates. Radii are given in arcsec.
The parameters are presented in
Table \ref{tabela2-new}  for the new 
objects and in Table \ref{tabela3-old} for the already known objects.

\section{The OGLE Collection of Star Clusters in the Outer Regions of the LMC}

We have found 226 new star clusters in the 225 square degrees area of
the outer regions of the LMC, based on observations
collected by the OGLE-IV survey. All those objects were numbered according to
the OGLE-IV naming scheme. The name consists of 4 parts: the first is
the name of the project (OGLE), the second is the abbreviated name of
the Large Magellanic Cloud (LMC), the third is the name of the type of the
object (CL for star CLusters) and the last part is the number of the
object in the OGLE-IV list. To make the numbering consistent with the OGLE-II
catalog (Pietrzy{\'n}ski \etal 1999), we start it at 0746.

Table \ref{tabela2-new} presents the OGLE collection of new star
clusters. Column 1 contains the OGLE identification number, column 2
shows the field name, in columns 3 and 4 we list the equatorial
coordinates (J2000) of cluster center, in column 5 the reliability
index and the last column contains the radius of the
cluster in arcseconds.

{\scriptsize
\begin{longtable}{|c|c|c|c|c|c|} 
\caption{New star clusters.
Columns contain: OGLE-IV name (1), OGLE-IV field name (2), centroids
($RA_{2000}$ (3) and $DEC_{2000}$ (4)), reliability index (5), radius (6) }
\label{tabela2-new} \\

\hline
\rule[0pt]{0pt}{2.9ex}OGLE-IV name & OGLE-IV field & RA (h:m:s) & DEC (d:m:s) & reliability & R$_{KDE}$ ["] \\ \hline
\endfirsthead

\multicolumn{6}{c}{{\small \tablename\ \thetable{} -- continued}} \\[2pt]
\hline
\rule[0pt]{0pt}{2.9ex}OGLE-IV name & OGLE-IV field & RA (h:m:s) & DEC (d:m:s) & reliability & R$_{KDE}$ ["] \\ \hline
\endhead

\hline
\endfoot

OGLE-LMC-CL-0746 & LMC600.16 & 5:45:02.72 & -63:48:43.0 & 1 & 49\\
OGLE-LMC-CL-0747 & LMC600.18 & 5:54:57.81 & -63:29:24.4 & 1 & 42\\
OGLE-LMC-CL-0748 & LMC600.22 & 5:49:49.14 & -63:36:01.6 & 0.9 & 49\\
OGLE-LMC-CL-0749 & LMC600.23 & 5:48:29.67 & -63:44:33.4 & 0.8 & 51\\
OGLE-LMC-CL-0750 & LMC601.14 & 5:47:21.41 & -62:49:41.2 & 0.6 & 55\\
OGLE-LMC-CL-0751 & LMC603.05 & 6:00:23.98 & -64:53:11.7 & 1 & 37\\
OGLE-LMC-CL-0752 & LMC603.10 & 6:05:15.46 & -64:33:20.1 & 0.6 & 45\\
OGLE-LMC-CL-0753 & LMC603.18 & 6:06:34.49 & -64:20:23.1 & 1 & 33\\
OGLE-LMC-CL-0754 & LMC603.25 & 5:56:16.88 & -64:14:25.6 & 0.8 & 47\\
OGLE-LMC-CL-0755 & LMC603.25 & 5:56:20.46 & -64:22:19.0 & 0.6 & 40\\
OGLE-LMC-CL-0756 & LMC605.16 & 5:53:37.44 & -62:12:17.2 & 0.7 & 50\\
OGLE-LMC-CL-0757 & LMC606.29 & 6:10:50.00 & -63:12:53.3 & 0.7 & 46\\
OGLE-LMC-CL-0758 & LMC607.01 & 6:17:42.12 & -65:32:53.3 & 1 & 24\\
OGLE-LMC-CL-0759 & LMC607.13 & 6:11:16.44 & -65:09:42.3 & 1 & 40\\
OGLE-LMC-CL-0760 & LMC608.07 & 6:10:23.58 & -66:43:03.7 & 1 & 47\\
OGLE-LMC-CL-0761 & LMC608.09 & 6:19:53.96 & -66:30:04.9 & 0.7 & 47\\
OGLE-LMC-CL-0762 & LMC608.16 & 6:10:02.25 & -66:21:39.9 & 1 & 37\\
OGLE-LMC-CL-0763 & LMC608.21 & 6:15:07.95 & -66:08:40.9 & 1 & 37\\
OGLE-LMC-CL-0764 & LMC608.21 & 6:15:13.38 & -65:57:11.6 & 0.8 & 43\\
OGLE-LMC-CL-0765 & LMC608.23 & 6:12:49.65 & -66:12:22.0 & 0.7 & 47\\
OGLE-LMC-CL-0766 & LMC608.24 & 6:11:30.80 & -66:12:40.7 & 0.6 & 47\\
OGLE-LMC-CL-0767 & LMC608.24 & 6:10:46.08 & -66:11:40.8 & 0.6 & 49\\
OGLE-LMC-CL-0768 & LMC608.24 & 6:11:15.15 & -65:59:22.8 & 0.6 & 46\\
OGLE-LMC-CL-0769 & LMC608.31 & 6:12:38.91 & -65:48:58.3 & 0.9 & 43\\
OGLE-LMC-CL-0770 & LMC609.05 & 6:16:28.75 & -67:47:50.5 & 1 & 39\\
OGLE-LMC-CL-0771 & LMC609.08 & 6:25:14.94 & -67:28:10.8 & 1 & 31\\
OGLE-LMC-CL-0772 & LMC609.14 & 6:15:20.57 & -67:34:06.7 & 1 & 51\\
OGLE-LMC-CL-0773 & LMC609.21 & 6:18:55.68 & -67:17:31.4 & 1 & 42\\
OGLE-LMC-CL-0774 & LMC609.29 & 6:18:15.65 & -66:52:12.1 & 0.9 & 39\\
OGLE-LMC-CL-0775 & LMC609.31 & 6:15:04.55 & -66:52:55.2 & 1 & 47\\
OGLE-LMC-CL-0776 & LMC610.02 & 6:24:22.99 & -69:06:26.2 & 1 & 43\\
OGLE-LMC-CL-0777 & LMC610.08 & 6:28:24.08 & -68:53:45.4 & 0.6 & 46\\
OGLE-LMC-CL-0778 & LMC610.23 & 6:17:40.42 & -68:28:05.4 & 0.6 & 46\\
OGLE-LMC-CL-0779 & LMC610.25 & 6:14:27.84 & -68:37:42.1 & 0.6 & 48\\
OGLE-LMC-CL-0780 & LMC610.29 & 6:21:03.26 & -68:15:35.2 & 0.9 & 47\\
OGLE-LMC-CL-0781 & LMC611.03 & 6:26:56.69 & -70:18:04.0 & 1 & 42\\
OGLE-LMC-CL-0782 & LMC611.04 & 6:24:57.34 & -70:15:07.6 & 1 & 35\\
OGLE-LMC-CL-0783 & LMC611.07 & 6:19:18.86 & -70:15:33.2 & 0.7 & 50\\
OGLE-LMC-CL-0784 & LMC611.15 & 6:19:11.14 & -70:09:19.8 & 0.6 & 38\\
OGLE-LMC-CL-0785 & LMC611.16 & 6:16:57.33 & -70:00:18.2 & 0.8 & 47\\
OGLE-LMC-CL-0786 & LMC612.03 & 6:29:33.86 & -71:41:40.8 & 0.7 & 24\\
OGLE-LMC-CL-0787 & LMC612.05 & 6:26:58.16 & -71:44:35.5 & 0.6 & 33\\
OGLE-LMC-CL-0788 & LMC612.05 & 6:27:06.17 & -71:42:35.5 & 0.6 & 35\\
OGLE-LMC-CL-0789 & LMC612.14 & 6:23:57.02 & -71:14:10.1 & 1 & 43\\
OGLE-LMC-CL-0790 & LMC612.14 & 6:24:50.63 & -71:26:14.7 & 0.7 & 46\\
OGLE-LMC-CL-0791 & LMC612.15 & 6:21:53.17 & -71:21:40.9 & 0.7 & 42\\
OGLE-LMC-CL-0792 & LMC613.31 & 6:28:44.91 & -71:57:21.8 & 1 & 47\\
OGLE-LMC-CL-0793 & LMC615.25 & 6:27:19.14 & -69:01:35.7 & 0.6 & 40\\
OGLE-LMC-CL-0794 & LMC631.09 & 4:51:24.24 & -75:36:42.1 & 0.7 & 32\\
OGLE-LMC-CL-0795 & LMC631.18 & 4:51:27.95 & -75:18:58.5 & 0.6 & 42\\
OGLE-LMC-CL-0796 & LMC631.19 & 4:49:34.08 & -75:21:54.7 & 0.6 & 32\\
OGLE-LMC-CL-0797 & LMC635.22 & 4:58:27.70 & -75:52:37.4 & 0.7 & 32\\
OGLE-LMC-CL-0798 & LMC635.22 & 4:59:39.50 & -76:02:51.5 & 0.6 & 29\\
OGLE-LMC-CL-0799 & LMC635.32 & 4:55:46.90 & -75:35:06.7 & 0.6 & 33\\
OGLE-LMC-CL-0800 & LMC636.04 & 5:01:00.08 & -75:11:50.5 & 0.8 & 41\\
OGLE-LMC-CL-0801 & LMC636.07 & 4:55:13.18 & -75:17:17.3 & 0.8 & 30\\
OGLE-LMC-CL-0802 & LMC636.07 & 4:55:11.75 & -75:12:25.2 & 0.7 & 45\\
OGLE-LMC-CL-0803 & LMC636.10 & 5:06:54.02 & -75:06:46.9 & 1 & 42\\
OGLE-LMC-CL-0804 & LMC636.14 & 4:56:57.88 & -74:52:32.7 & 0.6 & 58\\
OGLE-LMC-CL-0805 & LMC636.18 & 5:09:28.41 & -74:32:41.0 & 0.7 & 37\\
OGLE-LMC-CL-0806 & LMC636.22 & 5:00:14.82 & -74:50:03.7 & 0.7 & 36\\
OGLE-LMC-CL-0807 & LMC640.08 & 5:30:41.26 & -75:40:05.5 & 1 & 28\\
OGLE-LMC-CL-0808 & LMC640.12 & 5:19:54.77 & -75:28:56.0 & 0.6 & 44\\
OGLE-LMC-CL-0809 & LMC640.13 & 5:17:07.92 & -75:31:56.8 & 0.6 & 32\\
OGLE-LMC-CL-0810 & LMC640.18 & 5:27:06.84 & -75:21:20.4 & 1 & 29\\
OGLE-LMC-CL-0811 & LMC640.18 & 5:27:39.18 & -75:19:42.4 & 1 & 31\\
OGLE-LMC-CL-0812 & LMC640.19 & 5:26:05.46 & -75:24:59.9 & 1 & 33\\
OGLE-LMC-CL-0813 & LMC640.24 & 5:13:08.73 & -75:12:27.2 & 0.6 & 41\\
OGLE-LMC-CL-0814 & LMC644.15 & 5:29:38.69 & -76:12:10.7 & 0.9 & 29\\
OGLE-LMC-CL-0815 & LMC644.27 & 5:42:06.30 & -75:32:37.1 & 1 & 24\\
OGLE-LMC-CL-0816 & LMC645.07 & 5:29:53.58 & -75:18:40.0 & 0.9 & 43\\
OGLE-LMC-CL-0817 & LMC645.15 & 5:29:56.13 & -75:05:09.8 & 0.6 & 32\\
OGLE-LMC-CL-0818 & LMC645.15 & 5:31:39.70 & -75:00:07.0 & 0.6 & 55\\
OGLE-LMC-CL-0819 & LMC645.26 & 5:45:34.46 & -74:19:32.8 & 1 & 33\\
OGLE-LMC-CL-0820 & LMC649.21 & 5:56:49.25 & -75:11:21.4 & 0.6 & 35\\
OGLE-LMC-CL-0821 & LMC649.23 & 5:52:09.37 & -75:14:51.8 & 0.6 & 32\\
OGLE-LMC-CL-0822 & LMC649.24 & 5:48:00.65 & -75:13:39.6 & 0.6 & 44\\
OGLE-LMC-CL-0823 & LMC649.29 & 5:56:56.90 & -74:52:43.9 & 0.6 & 39\\
OGLE-LMC-CL-0824 & LMC535.11 & 4:32:06.04 & -73:40:10.0 & 0.6 & 45\\
OGLE-LMC-CL-0825 & LMC535.17 & 4:38:29.91 & -73:21:41.0 & 0.6 & 40\\
OGLE-LMC-CL-0826 & LMC535.19 & 4:35:35.82 & -73:25:07.0 & 0.9 & 30\\
OGLE-LMC-CL-0827 & LMC535.28 & 4:32:59.51 & -73:13:42.5 & 1 & 31\\
OGLE-LMC-CL-0828 & LMC536.28 & 4:34:37.88 & -71:54:17.7 & 0.9 & 40\\
OGLE-LMC-CL-0829 & LMC545.08 & 4:32:26.92 & -70:44:44.4 & 0.9 & 48\\
OGLE-LMC-CL-0830 & LMC546.21 & 4:27:42.08 & -69:07:51.5 & 0.6 & 36\\
OGLE-LMC-CL-0831 & LMC547.08 & 4:37:04.19 & -68:15:14.1 & 1 & 44\\
OGLE-LMC-CL-0832 & LMC547.08 & 4:36:30.12 & -68:06:26.4 & 0.9 & 49\\
OGLE-LMC-CL-0833 & LMC547.17 & 4:36:36.69 & -67:58:41.6 & 0.9 & 51\\
OGLE-LMC-CL-0834 & LMC547.20 & 4:31:17.54 & -67:46:50.5 & 1 & 32\\
OGLE-LMC-CL-0835 & LMC541.04 & 4:44:48.00 & -66:45:51.0 & 1 & 44\\
OGLE-LMC-CL-0836 & LMC541.06 & 4:41:42.23 & -66:50:54.3 & 0.6 & 36\\
OGLE-LMC-CL-0837 & LMC541.13 & 4:43:00.09 & -66:22:58.9 & 0.6 & 39\\
OGLE-LMC-CL-0838 & LMC541.17 & 4:49:51.76 & -66:08:36.3 & 0.6 & 46\\
OGLE-LMC-CL-0839 & LMC541.27 & 4:47:06.30 & -65:47:23.6 & 1 & 32\\
OGLE-LMC-CL-0840 & LMC541.27 & 4:46:41.09 & -65:53:13.0 & 0.6 & 43\\
OGLE-LMC-CL-0841 & LMC548.06 & 4:29:25.88 & -67:26:45.6 & 0.9 & 32\\
OGLE-LMC-CL-0842 & LMC542.04 & 4:45:32.70 & -65:29:58.0 & 1 & 36\\
OGLE-LMC-CL-0843 & LMC542.26 & 4:49:50.96 & -64:36:39.1 & 0.6 & 42\\
OGLE-LMC-CL-0844 & LMC586.26 & 4:50:59.83 & -63:16:52.0 & 0.70.9 & 33\\
OGLE-LMC-CL-0845 & LMC565.04 & 6:20:15.02 & -73:24:53.9 & 1 & 26\\
OGLE-LMC-CL-0846 & LMC565.07 & 6:14:19.67 & -73:19:24.0 & 0.9 & 32\\
OGLE-LMC-CL-0847 & LMC565.22 & 6:17:45.20 & -72:42:22.7 & 1 & 36\\
OGLE-LMC-CL-0848 & LMC565.24 & 6:14:40.96 & -72:53:39.5 & 1 & 32\\
OGLE-LMC-CL-0849 & LMC577.30 & 6:23:25.97 & -73:49:40.7 & 1 & 27\\
OGLE-LMC-CL-0850 & LMC597.03 & 5:41:58.33 & -63:31:13.0 & 0.6 & 47\\
OGLE-LMC-CL-0851 & LMC597.15 & 5:35:30.96 & -63:22:49.3 & 0.8 & 44\\
OGLE-LMC-CL-0852 & LMC597.32 & 5:36:08.37 & -62:44:15.4 & 0.6 & 43\\
OGLE-LMC-CL-0853 & LMC597.32 & 5:36:35.37 & -62:44:09.8 & 0.9 & 59\\
OGLE-LMC-CL-0854 & LMC598.02 & 5:42:05.20 & -62:26:50.7 & 0.6 & 43\\
OGLE-LMC-CL-0855 & LMC598.05 & 5:37:51.78 & -62:16:32.1 & 0.9 & 31\\
OGLE-LMC-CL-0856 & LMC598.05 & 5:37:40.44 & -62:29:07.8 & 0.6 & 45\\
OGLE-LMC-CL-0857 & LMC598.06 & 5:36:37.51 & -62:28:06.1 & 0.7 & 45\\
OGLE-LMC-CL-0858 & LMC598.15 & 5:34:41.77 & -62:02:55.0 & 0.6 & 49\\
OGLE-LMC-CL-0859 & LMC598.23 & 5:36:10.11 & -61:54:32.0 & 0.9 & 25\\
OGLE-LMC-CL-0860 & LMC593.11 & 5:30:30.64 & -62:46:34.1 & 1 & 42\\
OGLE-LMC-CL-0861 & LMC593.14 & 5:26:16.19 & -62:43:52.1 & 0.6 & 41\\
OGLE-LMC-CL-0862 & LMC593.26 & 5:32:56.59 & -62:07:47.0 & 0.9 & 42\\
OGLE-LMC-CL-0863 & LMC594.05 & 5:28:22.38 & -61:43:18.9 & 1 & 37\\
OGLE-LMC-CL-0864 & LMC576.05 & 5:17:52.07 & -61:12:32.1 & 0.7 & 33\\
OGLE-LMC-CL-0865 & LMC575.03 & 5:20:07.04 & -62:24:41.6 & 0.6 & 40\\
OGLE-LMC-CL-0866 & LMC575.05 & 5:17:45.46 & -62:28:40.7 & 0.7 & 39\\
OGLE-LMC-CL-0867 & LMC575.06 & 5:16:17.94 & -62:14:40.2 & 1 & 27\\
OGLE-LMC-CL-0868 & LMC574.01 & 5:23:46.89 & -63:37:38.5 & 1 & 34\\
OGLE-LMC-CL-0869 & LMC574.12 & 5:18:50.59 & -63:17:03.2 & 1 & 53\\
OGLE-LMC-CL-0870 & LMC574.18 & 5:23:25.39 & -62:53:24.4 & 1 & 43\\
OGLE-LMC-CL-0871 & LMC574.27 & 5:21:40.36 & -62:47:24.4 & 0.9 & 31\\
OGLE-LMC-CL-0872 & LMC574.27 & 5:22:11.68 & -62:44:44.0 & 0.6 & 47\\
OGLE-LMC-CL-0873 & LMC579.02 & 5:10:54.36 & -62:52:17.3 & 1 & 32\\
OGLE-LMC-CL-0874 & LMC579.11 & 5:10:00.59 & -62:34:55.1 & 1 & 22\\
OGLE-LMC-CL-0875 & LMC578.28 & 5:10:03.89 & -63:19:33.6 & 0.6 & 46\\
OGLE-LMC-CL-0876 & LMC578.31 & 5:05:57.81 & -63:25:06.3 & 1 & 38\\
OGLE-LMC-CL-0877 & LMC582.01 & 5:01:07.61 & -64:52:54.9 & 1 & 38\\
OGLE-LMC-CL-0878 & LMC582.02 & 5:00:05.61 & -64:43:23.8 & 0.9 & 52\\
OGLE-LMC-CL-0879 & LMC582.04 & 4:56:27.82 & -64:55:51.1 & 0.9 & 47\\
OGLE-LMC-CL-0880 & LMC582.12 & 4:57:34.04 & -64:32:34.0 & 0.6 & 37\\
OGLE-LMC-CL-0881 & LMC583.01 & 5:01:37.76 & -63:40:03.8 & 0.9 & 37\\
OGLE-LMC-CL-0882 & LMC583.18 & 5:01:43.71 & -63:07:41.8 & 0.9 & 32\\
OGLE-LMC-CL-0883 & LMC610.07 & 6:15:51.71 & -69:11:04.8 & 1 & 43\\
OGLE-LMC-CL-0884 & LMC601.09 & 5:54:10.41 & -62:42:17.9 & 1 & 31\\
OGLE-LMC-CL-0885 & LMC601.04 & 5:49:30.75 & -62:52:22.9 & 0.7 & 36\\
OGLE-LMC-CL-0886 & LMC631.17 & 4:52:44.28 & -75:16:30.6 & 1 & 29\\
OGLE-LMC-CL-0887 & LMC564.01 & 5:56:16.39 & -65:36:46.1 & 1 & 46\\
OGLE-LMC-CL-0888 & LMC564.02 & 5:54:57.66 & -65:20:52.7 & 1 & 50\\
OGLE-LMC-CL-0889 & LMC564.04 & 5:52:34.20 & -65:32:46.1 & 1 & 47\\
OGLE-LMC-CL-0890 & LMC564.04 & 5:53:00.66 & -65:25:51.3 & 0.7 & 43\\
OGLE-LMC-CL-0891 & LMC564.04 & 5:52:25.98 & -65:24:56.7 & 0.9 & 50\\
OGLE-LMC-CL-0892 & LMC564.04 & 5:52:34.55 & -65:29:37.9 & 0.8 & 50\\
OGLE-LMC-CL-0893 & LMC564.08 & 5:58:00.89 & -65:17:02.9 & 0.9 & 48\\
OGLE-LMC-CL-0894 & LMC564.08 & 5:57:47.59 & -65:01:19.6 & 0.9 & 50\\
OGLE-LMC-CL-0895 & LMC564.10 & 5:55:06.70 & -65:03:17.8 & 1 & 48\\
OGLE-LMC-CL-0896 & LMC564.11 & 5:53:11.95 & -65:01:40.7 & 0.9 & 48\\
OGLE-LMC-CL-0897 & LMC564.11 & 5:54:29.98 & -65:15:15.9 & 1 & 26\\
OGLE-LMC-CL-0898 & LMC564.11 & 5:53:48.08 & -65:15:47.1 & 0.9 & 45\\
OGLE-LMC-CL-0899 & LMC564.19 & 5:54:45.11 & -64:52:16.8 & 0.9 & 46\\
OGLE-LMC-CL-0900 & LMC596.06 & 5:37:33.68 & -64:52:07.8 & 0.8 & 45\\
OGLE-LMC-CL-0901 & LMC596.15 & 5:36:19.51 & -64:39:25.2 & 0.8 & 48\\
OGLE-LMC-CL-0902 & LMC596.16 & 5:34:31.55 & -64:39:11.0 & 0.8 & 59\\
OGLE-LMC-CL-0903 & LMC596.19 & 5:43:39.13 & -64:07:43.3 & 0.9 & 46\\
OGLE-LMC-CL-0904 & LMC596.24 & 5:36:15.90 & -64:08:48.5 & 1 & 49\\
OGLE-LMC-CL-0905 & LMC596.28 & 5:42:38.15 & -63:51:16.2 & 0.9 & 50\\
OGLE-LMC-CL-0906 & LMC596.30 & 5:39:58.90 & -63:59:32.5 & 0.9 & 52\\
OGLE-LMC-CL-0907 & LMC592.01 & 5:34:13.71 & -64:15:24.7 & 1 & 48\\
OGLE-LMC-CL-0908 & LMC592.01 & 5:34:22.73 & -64:09:23.7 & 1 & 52\\
OGLE-LMC-CL-0909 & LMC592.01 & 5:34:18.66 & -64:08:32.3 & 1 & 48\\
OGLE-LMC-CL-0910 & LMC592.03 & 5:31:59.53 & -64:06:51.6 & 0.9 & 21\\
OGLE-LMC-CL-0911 & LMC592.03 & 5:31:22.85 & -64:22:29.8 & 0.9 & 50\\
OGLE-LMC-CL-0912 & LMC592.05 & 5:28:23.29 & -64:19:45.7 & 0.9 & 48\\
OGLE-LMC-CL-0913 & LMC592.06 & 5:27:14.86 & -64:21:15.2 & 1 & 42\\
OGLE-LMC-CL-0914 & LMC592.06 & 5:27:39.25 & -64:16:11.6 & 0.9 & 49\\
OGLE-LMC-CL-0915 & LMC592.07 & 5:26:02.07 & -64:18:44.0 & 1 & 46\\
OGLE-LMC-CL-0916 & LMC592.08 & 5:35:33.68 & -63:57:18.9 & 0.9 & 50\\
OGLE-LMC-CL-0917 & LMC592.10 & 5:32:40.68 & -64:01:20.8 & 0.8 & 53\\
OGLE-LMC-CL-0918 & LMC592.11 & 5:31:15.42 & -64:01:38.8 & 0.9 & 45\\
OGLE-LMC-CL-0919 & LMC592.18 & 5:34:18.32 & -63:35:03.9 & 0.8 & 49\\
OGLE-LMC-CL-0920 & LMC592.21 & 5:30:13.04 & -63:38:01.6 & 0.8 & 53\\
OGLE-LMC-CL-0921 & LMC592.22 & 5:28:17.48 & -63:42:45.4 & 0.8 & 44\\
OGLE-LMC-CL-0922 & LMC592.22 & 5:29:07.14 & -63:41:46.9 & 0.7 & 37\\
OGLE-LMC-CL-0923 & LMC592.23 & 5:27:03.99 & -63:33:20.8 & 1 & 46\\
OGLE-LMC-CL-0924 & LMC592.28 & 5:31:34.57 & -63:14:00.6 & 0.8 & 42\\
OGLE-LMC-CL-0925 & LMC592.30 & 5:28:22.61 & -63:26:18.1 & 0.8 & 42\\
OGLE-LMC-CL-0926 & LMC573.01 & 5:23:42.86 & -64:55:57.0 & 0.7 & 49\\
OGLE-LMC-CL-0927 & LMC573.03 & 5:20:38.30 & -64:58:10.1 & 0.7 & 46\\
OGLE-LMC-CL-0928 & LMC573.03 & 5:19:54.87 & -64:43:10.2 & 0.7 & 45\\
OGLE-LMC-CL-0929 & LMC573.04 & 5:18:36.26 & -64:46:37.2 & 0.9 & 52\\
OGLE-LMC-CL-0930 & LMC573.06 & 5:15:53.62 & -64:59:29.0 & 1 & 53\\
OGLE-LMC-CL-0931 & LMC573.06 & 5:15:56.93 & -64:57:49.5 & 0.8 & 52\\
OGLE-LMC-CL-0932 & LMC573.06 & 5:15:30.51 & -64:42:36.1 & 1 & 41\\
OGLE-LMC-CL-0933 & LMC573.07 & 5:14:54.26 & -64:50:58.7 & 0.7 & 51\\
OGLE-LMC-CL-0934 & LMC573.08 & 5:24:43.39 & -64:23:38.1 & 0.7 & 54\\
OGLE-LMC-CL-0935 & LMC573.09 & 5:23:52.59 & -64:39:42.9 & 0.6 & 43\\
OGLE-LMC-CL-0936 & LMC573.12 & 5:19:17.36 & -64:34:44.0 & 0.6 & 44\\
OGLE-LMC-CL-0937 & LMC573.21 & 5:18:53.57 & -64:21:35.5 & 0.8 & 50\\
OGLE-LMC-CL-0938 & LMC573.29 & 5:18:55.59 & -63:55:05.0 & 0.9 & 48\\
OGLE-LMC-CL-0939 & LMC573.31 & 5:16:32.55 & -63:58:09.1 & 0.7 & 44\\
OGLE-LMC-CL-0940 & LMC513.02 & 5:10:00.36 & -65:30:42.5 & 0.8 & 46\\
OGLE-LMC-CL-0941 & LMC513.02 & 5:11:10.01 & -65:26:59.2 & 0.7 & 54\\
OGLE-LMC-CL-0942 & LMC513.04 & 5:07:46.80 & -65:29:27.3 & 0.8 & 49\\
OGLE-LMC-CL-0943 & LMC513.09 & 5:11:45.04 & -65:12:08.0 & 1 & 47\\
OGLE-LMC-CL-0944 & LMC513.09 & 5:12:28.65 & -65:13:49.7 & 0.8 & 54\\
OGLE-LMC-CL-0945 & LMC513.11 & 5:09:51.54 & -65:07:52.1 & 0.8 & 44\\
OGLE-LMC-CL-0946 & LMC513.11 & 5:09:21.20 & -65:17:38.8 & 0.6 & 48\\
OGLE-LMC-CL-0947 & LMC513.16 & 5:02:17.33 & -65:04:55.7 & 0.7 & 50\\
OGLE-LMC-CL-0948 & LMC513.18 & 5:12:03.60 & -64:56:46.5 & 0.7 & 50\\
OGLE-LMC-CL-0949 & LMC520.02 & 5:33:13.14 & -65:29:37.2 & 0.7 & 52\\
OGLE-LMC-CL-0950 & LMC520.03 & 5:31:27.20 & -65:20:55.8 & 0.7 & 53\\
OGLE-LMC-CL-0951 & LMC520.03 & 5:32:04.58 & -65:20:35.6 & 1 & 56\\
OGLE-LMC-CL-0952 & LMC520.04 & 5:30:30.99 & -65:22:17.2 & 0.7 & 52\\
OGLE-LMC-CL-0953 & LMC520.04 & 5:30:42.80 & -65:23:42.4 & 0.6 & 52\\
OGLE-LMC-CL-0954 & LMC520.05 & 5:28:20.22 & -65:29:15.1 & 0.6 & 50\\
OGLE-LMC-CL-0955 & LMC520.05 & 5:28:56.72 & -65:29:54.4 & 0.9 & 53\\
OGLE-LMC-CL-0956 & LMC520.07 & 5:26:06.91 & -65:19:56.5 & 0.7 & 49\\
OGLE-LMC-CL-0957 & LMC520.09 & 5:34:21.84 & -65:03:35.4 & 0.8 & 49\\
OGLE-LMC-CL-0958 & LMC520.10 & 5:33:43.75 & -65:10:44.4 & 0.7 & 54\\
OGLE-LMC-CL-0959 & LMC520.13 & 5:28:17.19 & -65:00:50.3 & 0.8 & 42\\
OGLE-LMC-CL-0960 & LMC520.15 & 5:26:02.62 & -65:10:14.2 & 0.8 & 51\\
OGLE-LMC-CL-0961 & LMC520.15 & 5:25:42.53 & -65:06:26.0 & 1 & 49\\
OGLE-LMC-CL-0962 & LMC520.16 & 5:23:49.33 & -65:08:52.4 & 0.9 & 45\\
OGLE-LMC-CL-0963 & LMC520.21 & 5:29:41.66 & -64:57:28.7 & 0.7 & 45\\
OGLE-LMC-CL-0964 & LMC520.24 & 5:26:36.81 & -64:46:53.8 & 0.8 & 46\\
OGLE-LMC-CL-0965 & LMC520.25 & 5:24:48.58 & -64:48:42.0 & 0.6 & 51\\
OGLE-LMC-CL-0966 & LMC520.25 & 5:24:29.97 & -64:50:04.3 & 1 & 57\\
OGLE-LMC-CL-0967 & LMC520.26 & 5:34:28.96 & -64:38:13.5 & 0.7 & 49\\
OGLE-LMC-CL-0968 & LMC520.28 & 5:31:44.22 & -64:30:55.1 & 0.8 & 50\\
OGLE-LMC-CL-0969 & LMC520.29 & 5:30:20.58 & -64:32:41.3 & 0.8 & 47\\
OGLE-LMC-CL-0970 & LMC520.30 & 5:28:31.47 & -64:34:12.9 & 0.7 & 60\\
OGLE-LMC-CL-0971 & LMC520.31 & 5:26:56.80 & -64:23:20.2 & 0.6 & 49\\
\hline
\end{longtable}
}

Table \ref{tabela3-old} presents the OGLE collection of star clusters,
which were already known. We  cross-matched our detections with the
Bica catalog and the DES catalog. All but six clusters listed in those
catalogs which are located in the analyzed area of the LMC were also
found in OGLE data. The names in column 2 were taken mostly from the
Bica catalog. The DES catalog (Pieres \etal 2016) contains only 28
objects which are cataloged by the authors as new ones. We identified
12 of them as listed in the Bica catalog as type AC (association similar 
to cluster) or A (association) objects. Thus we
listed only 15 clusters as genuine DES new findings. DES001SC12 is
located in the gap between OGLE-IV subfields so our algorithm did not detected
it. Nevertheless, we include the DES names for all 27 clusters in column 3.

Column 1 contains the OGLE identification number, column 2 shows the
field name. In column 3 we list the cross-identification of clusters,
in column 4 is cluster type (C -- ordinary cluster, CN --
cluster in nebula, CA -- cluster similar to association, A -- ordinary
association, AC -- association similar to cluster). Columns 5 and 6
show our estimation of the equatorial coordinates (J2000) and column 7
shows our estimation of the radius in arcseconds. 
For all
previously known clusters we calculated the reliability parameter equal
to one, so we do not list their values here.

{\tiny
\begin{center}
\begin{longtable}{|c|c|c|c|c|c|c|}
\caption{Already known star clusters.
Columns contain: OGLE-IV name (1), OGLE-IV field name (2), previously
assigned names (3), types (4), centroids ($RA_{2000}$ (5) and
$DEC_{2000}$ (6)), radius (7) } \label{tabela3-old} \\

\hline
\rule[0pt]{0pt}{2.9ex}{\scriptsize OGLE-IV name} &
{\scriptsize OGLE-IV field} &
{\scriptsize name} &
{\scriptsize type from Bica} &
{\scriptsize RA (h:m:s)} &
{\scriptsize DEC (deg:m:s)} &
{\scriptsize $R_{KDE}$ ["]} \\ \hline
\endfirsthead

\multicolumn{7}{c}{{\small \tablename\ \thetable{} -- continued}} \\[2pt]
\hline
\rule[0pt]{0pt}{2.9ex}{\scriptsize OGLE-IV name} &
{\scriptsize OGLE-IV field} &
{\scriptsize name} &
{\scriptsize type from Bica} &
{\scriptsize RA (h:m:s)} &
{\scriptsize DEC (deg:m:s)} &
{\scriptsize $R_{KDE}$ ["]} \\ \hline
\endhead

\hline
\endfoot

OGLE-LMC-CL-0972 & LMC600.04 & LW319, KMHK1439 & C & 5:51:05.84 & -64:11:13.3 & 41\\
OGLE-LMC-CL-0973 & LMC600.05 & KMHK1391 & C & 5:49:02.12 & -64:20:53.6 & 39\\
OGLE-LMC-CL-0974 & LMC600.05 & SL738, LW314, KMHK1417 & C & 5:50:05.93 & -64:09:16.9 & 47\\
OGLE-LMC-CL-0975 & LMC600.10 & KMHK1484 & CA & 5:53:12.08 & -63:48:27.4 & 42\\
OGLE-LMC-CL-0976 & LMC600.11 & LW323, KMHK1455 & C & 5:51:45.50 & -63:51:19.5 & 33\\
OGLE-LMC-CL-0977 & LMC600.15 & SL710, LW298, ESO86SC32, KMHK1356 & C & 5:47:02.17 & -63:52:37.5 & 45\\
OGLE-LMC-CL-0978 & LMC600.15 & BSDL2976 & C & 5:46:50.44 & -63:51:24.3 & 49\\
OGLE-LMC-CL-0979 & LMC600.17 & KMHK1530 & CA & 5:56:08.30 & -63:38:34.6 & 37\\
OGLE-LMC-CL-0980 & LMC600.19 & SL768, LW326, ESO86SC38, KMHK1493 & C & 5:53:52.36 & -63:36:50.9 & 47\\
OGLE-LMC-CL-0981 & LMC600.20 & OHSC27, KMHK1469/BSDL3145 & C/CA & 5:52:29.93 & -63:36:03.5 & 34\\
OGLE-LMC-CL-0982 & LMC600.21 & NGC2120, SL742, LW316, ESO86SC34, & C & 5:50:35.23 & -63:40:58.8 & 45\\
OGLE-LMC-CL-0983 & LMC600.23 & KMHK1381 & C & 5:48:22.30 & -63:36:01.3 & 48\\
OGLE-LMC-CL-0984 & LMC600.25 & SL694, LW287, KMHK1318 & C & 5:45:06.50 & -63:40:36.3 & 31\\
OGLE-LMC-CL-0985 & LMC600.25 & SL701, LW289, ESO56SC30, KMHK1330 & C & 5:45:33.69 & -63:42:57.8 & 43\\
OGLE-LMC-CL-0986 & LMC600.31 & BSDL3032 & CA & 5:48:33.92 & -63:14:10.7 & 41\\
OGLE-LMC-CL-0987 & LMC601.03 & LW320, KMHK1447 & C & 5:51:21.50 & -62:59:38.5 & 30\\
OGLE-LMC-CL-0988 & LMC601.04 & SL729, LW311, KMHK1399 & C & 5:49:20.58 & -63:07:45.0 & 27\\
OGLE-LMC-CL-0989 & LMC601.14 & BSDL2991 & CA & 5:47:31.86 & -62:38:43.7 & 40\\
OGLE-LMC-CL-0990 & LMC601.16 & NGC2097, SL682, LW282, ESO56SC28, & C & 5:44:15.84 & -62:46:58.8 & 43\\
OGLE-LMC-CL-0991 & LMC601.17 & OHSC28 & C & 5:54:24.73 & -62:17:43.3 & 20\\
OGLE-LMC-CL-0992 & LMC603.03 & KMHK1615 & C & 6:03:43.94 & -64:43:58.5 & 33\\
OGLE-LMC-CL-0993 & LMC603.03 & DES001SC28 & - & 6:03:18.77 & -64:49:48.9 & 38\\
OGLE-LMC-CL-0994 & LMC603.05 & LW354, KMHK1581 & CA & 6:00:37.97 & -64:45:01.8 & 40\\
OGLE-LMC-CL-0995 & LMC603.07 & LW340, KMHK1548 & C & 5:57:35.01 & -64:59:48.1 & 34\\
OGLE-LMC-CL-0996 & LMC603.14 & DES001SC27 & - & 5:58:26.09 & -64:34:44.4 & 33\\
OGLE-LMC-CL-0997 & LMC603.21 & SL823, LW362, ESO86SC51, KMHK1596 & C & 6:02:18.58 & -64:19:53.6 & 32\\
OGLE-LMC-CL-0998 & LMC603.21 & KMHK1593 & C & 6:01:52.06 & -64:07:59.4 & 26\\
OGLE-LMC-CL-0999 & LMC603.32 & SL798, LW344, KMHK1556 & C & 5:58:12.14 & -63:53:41.4 & 34\\
OGLE-LMC-CL-1000 & LMC604.04 & NGC2162, SL814, LW351, ESO86SC47, & C & 6:00:27.75 & -63:43:09.6 & 44\\
OGLE-LMC-CL-1001 & LMC604.07 & KMHK1530 & CA & 5:56:08.40 & -63:38:34.0 & 30\\
OGLE-LMC-CL-1002 & LMC607.14 & SL855, LW420, ESO86SC66, KMHK1674 & C & 6:10:54.94 & -65:02:33.0 & 40\\
OGLE-LMC-CL-1003 & LMC607.19 & LW444, KMHK1710 & CA & 6:15:57.80 & -64:58:25.5 & 27\\
OGLE-LMC-CL-1004 & LMC607.24 & SL845, LW401, KMHK1656 & C & 6:08:38.26 & -64:48:08.8 & 30\\
OGLE-LMC-CL-1005 & LMC607.30 & KMHK1678 & C & 6:11:22.12 & -64:40:43.5 & 21\\
OGLE-LMC-CL-1006 & LMC608.13 & KMHK1699 & C & 6:14:40.85 & -66:25:16.6 & 28\\
OGLE-LMC-CL-1007 & LMC608.13 & LW436, KMHK1697 & C & 6:14:25.72 & -66:26:37.6 & 44\\
OGLE-LMC-CL-1008 & LMC608.22 & SL866, LW438, ESO87SC02, KMHK1698 & C & 6:14:30.90 & -65:58:43.7 & 42\\
OGLE-LMC-CL-1009 & LMC608.25 & SL850, LW413, ESO86SC64, KMHK1671 & C & 6:10:12.01 & -65:57:34.1 & 40\\
OGLE-LMC-CL-1010 & LMC609.03 & BSDL3245 & AC & 6:20:31.53 & -67:56:44.6 & 38\\
OGLE-LMC-CL-1011 & LMC609.03 & BM278 & AC & 6:20:49.16 & -68:00:46.1 & 39\\
OGLE-LMC-CL-1012 & LMC609.04 & LW454, BM272, KMHK1723 & C & 6:18:11.35 & -68:04:24.5 & 41\\
OGLE-LMC-CL-1013 & LMC609.06 & SL871, LW443, BM267, KMHK1708 & C & 6:15:25.64 & -67:59:01.0 & 33\\
OGLE-LMC-CL-1014 & LMC609.07 & SL863, LW433, BM263, KMHK1693 & C & 6:13:59.44 & -67:58:57.5 & 40\\
OGLE-LMC-CL-1015 & LMC609.07 & SL864, LW432, BM262, KMHK1694 & C & 6:13:56.19 & -68:04:58.6 & 48\\
OGLE-LMC-CL-1016 & LMC609.10 & SL885, LW467, BM279, KMHK1738 & C & 6:21:12.53 & -67:28:35.8 & 39\\
OGLE-LMC-CL-1017 & LMC609.11 & NGC2231, SL884, LW466, ESO87SC06, & C & 6:20:42.31 & -67:31:20.3 & 47\\
OGLE-LMC-CL-1018 & LMC609.12 & SL876, LW452, BM271, KMHK1721 & C & 6:18:08.31 & -67:29:07.8 & 40\\
OGLE-LMC-CL-1019 & LMC609.12 & LW458, BM273, KMHK1725 & C & 6:19:09.96 & -67:29:43.7 & 34\\
OGLE-LMC-CL-1020 & LMC609.15 & BM264, KMHK1695 & CA & 6:14:04.98 & -67:36:21.3 & 40\\
OGLE-LMC-CL-1021 & LMC609.15 & BM258 & CA & 6:12:50.88 & -67:39:21.1 & 38\\
OGLE-LMC-CL-1022 & LMC610.04 & SL886, LW468, KMHK1740 & C & 6:21:26.79 & -69:18:00.8 & 39\\
OGLE-LMC-CL-1023 & LMC610.05 & LW465 & AC & 6:20:19.73 & -69:04:27.6 & 40\\
OGLE-LMC-CL-1024 & LMC610.06 & LW451, KMHK1720 & C & 6:17:44.01 & -69:13:25.5 & 42\\
OGLE-LMC-CL-1025 & LMC610.07 & SL873, LW447, KMHK1715 & C & 6:16:15.51 & -69:07:34.7 & 31\\
OGLE-LMC-CL-1026 & LMC610.09 & NGC2249, SL893, LW479, ESO57SC82 & C & 6:25:53.14 & -68:55:20.4 & 43\\
OGLE-LMC-CL-1027 & LMC610.09 & BSDL3246 & A & 6:26:42.21 & -68:48:36.3 & 41\\
OGLE-LMC-CL-1028 & LMC610.11 & NGC2241, SL888, LW471, ESO57SC79 & C & 6:22:50.18 & -68:55:42.4 & 42\\
OGLE-LMC-CL-1029 & LMC610.11 & SL889, LW474, KMHK1748 & C & 6:23:31.98 & -68:59:05.3 & 39\\
OGLE-LMC-CL-1030 & LMC610.13 & BSDL3242 & A & 6:19:31.57 & -68:57:44.5 & 40\\
OGLE-LMC-CL-1031 & LMC610.15 & LW449, KMHK1717 & CA & 6:16:48.68 & -68:51:02.1 & 44\\
OGLE-LMC-CL-1032 & LMC610.20 & BM281, KMHK1747 & CA & 6:23:37.22 & -68:23:46.7 & 42\\
OGLE-LMC-CL-1033 & LMC610.24 & BM270, KMHK1718 & AC & 6:17:04.63 & -68:23:51.9 & 28\\
OGLE-LMC-CL-1034 & LMC610.28 & LW472, BM280, KMHK1746 & C & 6:23:11.59 & -68:19:13.6 & 39\\
OGLE-LMC-CL-1035 & LMC610.30 & LW459, BM274, KMHK1727 & C & 6:19:17.76 & -68:19:44.2 & 30\\
OGLE-LMC-CL-1036 & LMC610.30 & SL883, LW461, BM275, KMHK1731 & C & 6:19:56.53 & -68:15:09.1 & 32\\
OGLE-LMC-CL-1037 & LMC610.30 & BM276, KMHK1736 & AC & 6:20:26.41 & -68:21:31.9 & 22\\
OGLE-LMC-CL-1038 & LMC610.31 & LW454, BM272, KMHK1723 & C & 6:18:11.82 & -68:04:30.2 & 37\\
OGLE-LMC-CL-1039 & LMC610.32 & LW448, BM269, KMHK1716 & AC & 6:16:35.68 & -68:07:14.8 & 38\\
OGLE-LMC-CL-1040 & LMC610.32 & BM268, KMHK1711 & AC & 6:15:39.86 & -68:20:10.4 & 38\\
OGLE-LMC-CL-1041 & LMC611.05 & LW475,KMHK1750 & CA & 6:23:31.37 & -70:32:17.5 & 22\\
OGLE-LMC-CL-1042 & LMC611.16 & KMHK1719 & C & 6:17:19.99 & -70:03:36.8 & 29\\
OGLE-LMC-CL-1043 & LMC611.24 & KMHK1732 & C & 6:19:48.88 & -69:47:24.8 & 33\\
OGLE-LMC-CL-1044 & LMC611.26 & SL896, LW480, KMHK1758 & C & 6:29:58.85 & -69:20:04.4 & 24\\
OGLE-LMC-CL-1045 & LMC611.31 & SL886, LW468, KMHK1740 & C & 6:21:26.85 & -69:18:24.4 & 38\\
OGLE-LMC-CL-1046 & LMC612.06 & SL891, LW477, ESO57SC81, KMHK1752 & C & 6:24:51.23 & -71:39:35.0 & 31\\
OGLE-LMC-CL-1047 & LMC612.07 & SL890, LW473, KMHK1749 & C & 6:23:02.97 & -71:41:15.0 & 28\\
OGLE-LMC-CL-1048 & LMC612.16 & LW463, KMHK1733 & CA & 6:19:50.03 & -71:18:40.6 & 38\\
OGLE-LMC-CL-1049 & LMC612.19 & SL897, LW483, KMHK1760 & C & 6:33:01.13 & -71:07:43.9 & 36\\
OGLE-LMC-CL-1050 & LMC612.23 & SL892, LW478, KMHK1753 & C & 6:25:17.20 & -71:06:10.0 & 24\\
OGLE-LMC-CL-1051 & LMC612.25 & KMHK1739 & CA & 6:21:00.95 & -71:02:14.3 & 41\\
OGLE-LMC-CL-1052 & LMC612.28 & OHSC36, KMHK1757 & CA & 6:29:42.61 & -70:35:20.3 & 33\\
OGLE-LMC-CL-1053 & LMC612.32 & LW475, KMHK1750 & CA & 6:23:23.61 & -70:33:13.2 & 29\\
OGLE-LMC-CL-1054 & LMC613.12 & LW482, KMHK1759 & C & 6:32:22.91 & -72:28:20.1 & 21\\
OGLE-LMC-CL-1055 & LMC613.12 & KMHK1761 & AC & 6:33:00.59 & -72:26:03.6 & 28\\
OGLE-LMC-CL-1056 & LMC614.24 & OHSC35, BM282, KMHK1754 & CA & 6:25:42.12 & -67:59:48.4 & 32\\
OGLE-LMC-CL-1057 & LMC615.15 & SL896, LW480, KMHK1758 & C & 6:29:58.51 & -69:20:01.5 & 23\\
OGLE-LMC-CL-1058 & LMC631.09 & SL61, LW79, ESO32SC20, KMHK178 & C & 4:50:48.19 & -75:31:48.2 & 40\\
OGLE-LMC-CL-1059 & LMC631.10 & SL53, LW73, KMHK160 & C & 4:49:53.61 & -75:37:41.7 & 27\\
OGLE-LMC-CL-1060 & LMC631.26 & SL74, LW82, ESO32SC21, KMHK210 & C & 4:52:01.12 & -74:50:41.2 & 32\\
OGLE-LMC-CL-1061 & LMC631.26 & SL80, LW88, KMHK220 & C & 4:52:11.39 & -74:53:14.5 & 48\\
OGLE-LMC-CL-1062 & LMC631.28 & SL36, LW60, KMHK97 & C & 4:46:09.12 & -74:53:17.9 & 24\\
OGLE-LMC-CL-1063 & LMC631.29 & SL29, LW50, KMHK81 & C & 4:45:13.24 & -75:06:57.2 & 20\\
OGLE-LMC-CL-1064 & LMC635.26 & SL295, LW153, KMHK639 & C & 5:10:11.48 & -75:32:37.3 & 26\\
OGLE-LMC-CL-1065 & LMC635.26 & LW155, KMHK647 & C & 5:10:32.79 & -75:39:45.5 & 25\\
OGLE-LMC-CL-1066 & LMC635.27 & SL248, KMHK578 & C & 5:06:50.00 & -75:40:56.4 & 25\\
OGLE-LMC-CL-1067 & LMC635.28 & OHSC7 & C & 5:04:32.69 & -75:40:50.1 & 28\\
OGLE-LMC-CL-1068 & LMC635.29 & OHSC5 & C & 5:01:53.88 & -75:28:07.9 & 25\\
OGLE-LMC-CL-1069 & LMC635.30 & KMHK453 & C & 5:00:13.72 & -75:45:04.4 & 20\\
OGLE-LMC-CL-1070 & LMC635.31 & LW106, KMHK399 & C & 4:58:06.45 & -75:32:24.9 & 24\\
OGLE-LMC-CL-1071 & LMC636.02 & LW137, KMHK558 & C & 5:06:02.69 & -75:25:04.7 & 33\\
OGLE-LMC-CL-1072 & LMC636.04 & OHSC5 & C & 5:01:50.28 & -75:27:48.9 & 38\\
OGLE-LMC-CL-1073 & LMC636.05 & LW110, KMHK414 & C & 4:58:45.93 & -75:13:07.2 & 34\\
OGLE-LMC-CL-1074 & LMC636.06 & OHSC3, KMHK362 & C & 4:56:36.49 & -75:14:31.4 & 20\\
OGLE-LMC-CL-1075 & LMC636.11 & OHSC6, KMHK532 & C & 5:04:41.00 & -75:03:13.1 & 26\\
OGLE-LMC-CL-1076 & LMC636.12 & SL192, LW121, KMHK489 & C & 5:02:26.41 & -74:51:50.7 & 33\\
OGLE-LMC-CL-1077 & LMC636.13 & OHSC4, KMHK426 & C & 4:59:14.78 & -75:07:57.4 & 22\\
OGLE-LMC-CL-1078 & LMC636.15 & KMHK343/LW95,KMHK344 & C/C & 4:55:57.82 & -75:08:21.1 & 16\\
OGLE-LMC-CL-1079 & LMC636.16 & SL84, LW89, ESO32SC22, KMHK232 & C & 4:52:48.49 & -75:04:28.4 & 29\\
OGLE-LMC-CL-1080 & LMC636.16 & SL74, LW82, ESO32SC21, KMHK210 & C & 4:52:01.88 & -74:50:44.0 & 28\\
OGLE-LMC-CL-1081 & LMC636.16 & SL80, LW88, KMHK220 & C & 4:52:22.62 & -74:53:22.1 & 36\\
OGLE-LMC-CL-1082 & LMC636.17 & KMHK651 & C & 5:11:08.51 & -74:32:41.7 & 32\\
OGLE-LMC-CL-1083 & LMC636.18 & LW147, KMHK604 & C & 5:08:34.69 & -74:43:44.8 & 30\\
OGLE-LMC-CL-1084 & LMC636.19 & LW141, KMHK585 & C & 5:07:33.68 & -74:38:05.5 & 26\\
OGLE-LMC-CL-1085 & LMC636.20 & LW132, KMHK542 & C & 5:05:14.38 & -74:43:34.9 & 32\\
OGLE-LMC-CL-1086 & LMC636.22 & KMHK447 & C & 5:00:15.44 & -74:45:11.5 & 33\\
OGLE-LMC-CL-1087 & LMC636.22 & SL166, LW111, KMHK419 & C & 4:59:15.45 & -74:39:17.8 & 39\\
OGLE-LMC-CL-1088 & LMC636.22 & KMHK417 & C & 4:59:04.75 & -74:47:39.8 & 39\\
OGLE-LMC-CL-1089 & LMC636.23 & LW107, KMHK396 & C & 4:58:15.13 & -74:42:19.4 & 36\\
OGLE-LMC-CL-1090 & LMC636.23 & H88-65 & C & 4:58:13.01 & -74:35:21.5 & 29\\
OGLE-LMC-CL-1091 & LMC636.24 & SL118, LW94, KMHK323 & C & 4:55:31.49 & -74:40:33.4 & 29\\
OGLE-LMC-CL-1092 & LMC636.25 & H88-15, OHSC2, KMHK238 & C & 4:53:10.26 & -74:40:53.1 & 25\\
OGLE-LMC-CL-1093 & LMC636.25 & SL74, LW82, ESO32SC21, KMHK210 & C & 4:52:07.26 & -74:49:10.7 & 48\\
OGLE-LMC-CL-1094 & LMC636.32 & NGC1777, SL121, LW96, ESO33SC01 & C & 4:55:47.27 & -74:17:15.1 & 38\\
OGLE-LMC-CL-1095 & LMC523.12 & SL28, LW47, ESO32SC19, KMHK71 & C & 4:44:41.65 & -74:15:32.5 & 40\\
OGLE-LMC-CL-1096 & LMC523.15 & LW15, KMHK25 & C & 4:38:25.95 & -74:27:50.4 & 35\\
OGLE-LMC-CL-1097 & LMC523.17 & H88-16, KMHK228 & C & 4:52:56.72 & -74:00:59.6 & 37\\
OGLE-LMC-CL-1098 & LMC523.20 & LW62, KMHK96 & C & 4:46:18.72 & -74:09:38.0 & 33\\
OGLE-LMC-CL-1099 & LMC523.23 & SL13, LW17,KMHK31 & C & 4:39:42.39 & -74:01:06.6 & 25\\
OGLE-LMC-CL-1100 & LMC523.26 & LW76, SL59e, KMHK157, BRHT24b & C & 4:50:26.49 & -73:38:49.8 & 26\\
OGLE-LMC-CL-1101 & LMC523.27 & LW75, SL59w, KMHK152, BRHT24a & C & 4:50:13.46 & -73:38:37.9 & 32\\
OGLE-LMC-CL-1102 & LMC523.27 & LW66, KMHK129 & C & 4:49:04.22 & -73:50:21.1 & 27\\
OGLE-LMC-CL-1103 & LMC523.30 & KMHK58 & C & 4:43:13.92 & -73:48:48.1 & 51\\
OGLE-LMC-CL-1104 & LMC639.27 & SL455, LW207, KMHK891 & C & 5:24:17.75 & -76:12:37.6 & 30\\
OGLE-LMC-CL-1105 & LMC639.30 & OHSC9, KMHK733 & C & 5:16:23.49 & -76:12:23.3 & 21\\
OGLE-LMC-CL-1106 & LMC640.04 & SL400, LW188, KMHK809 & C & 5:19:37.51 & -75:48:35.5 & 26\\
OGLE-LMC-CL-1107 & LMC640.10 & SL451, LW206, KMHK883 & C & 5:24:13.45 & -75:33:56.8 & 28\\
OGLE-LMC-CL-1108 & LMC640.16 & SL295, LW153, KMHK639 & C & 5:10:10.86 & -75:32:37.8 & 25\\
OGLE-LMC-CL-1109 & LMC640.16 & LW155, KMHK647 & C & 5:10:32.90 & -75:39:42.7 & 22\\
OGLE-LMC-CL-1110 & LMC640.17 & LW231, KMHK1031 & C & 5:30:26.92 & -75:20:59.1 & 32\\
OGLE-LMC-CL-1111 & LMC640.20 & IC2134, SL437, LW198, ESO33SC19, & C & 5:23:07.63 & -75:26:43.6 & 30\\
OGLE-LMC-CL-1112 & LMC640.27 & OHSC16, KMHK881 & C & 5:24:22.45 & -74:50:31.8 & 30\\
OGLE-LMC-CL-1113 & LMC640.29 & OHSC11, KMHK803 & C & 5:19:36.02 & -75:06:05.1 & 31\\
OGLE-LMC-CL-1114 & LMC640.31 & OHSC8, KMHK708 & C & 5:15:18.30 & -74:56:07.3 & 29\\
OGLE-LMC-CL-1115 & LMC644.22 & SL603, LW249, KMHK1151 & C & 5:35:44.39 & -75:57:34.9 & 21\\
OGLE-LMC-CL-1116 & LMC644.29 & IC2148, SL642, LW265, ESO33SC28 & C & 5:39:12.14 & -75:33:43.6 & 25\\
OGLE-LMC-CL-1117 & LMC645.03 & SL647, LW267, KMHK1223 & C & 5:39:35.64 & -75:12:30.8 & 28\\
OGLE-LMC-CL-1118 & LMC645.03 & LW270, KMHK1243 & C & 5:40:23.60 & -75:18:07.4 & 40\\
OGLE-LMC-CL-1119 & LMC645.06 & IC2140, SL581, LW241, ESO33SC24 & C & 5:33:22.03 & -75:22:50.7 & 53\\
OGLE-LMC-CL-1120 & LMC645.07 & LW231, KMHK1031 & C & 5:30:25.72 & -75:20:58.6 & 40\\
OGLE-LMC-CL-1121 & LMC645.09 & SL703,LW290,KMHK1352 & C & 5:44:52.46 & -74:51:17.9 & 40\\
OGLE-LMC-CL-1122 & LMC645.12 & LW263,KMHK1208 & C & 5:39:08.40 & -74:51:14.7 & 36\\
OGLE-LMC-CL-1123 & LMC645.18 & LW297,KMHK1372 & C & 5:46:03.81 & -74:32:04.8 & 24\\
OGLE-LMC-CL-1124 & LMC645.21 & IC2146,SL632,LW258,ESO33SC26, & C & 5:37:42.96 & -74:46:43.8 & 47\\
OGLE-LMC-CL-1125 & LMC645.24 & LW234,KMHK1042 & C & 5:31:00.61 & -74:40:16.8 & 31\\
OGLE-LMC-CL-1126 & LMC645.26 & OHSC25,KMHK1338 & C & 5:44:05.28 & -74:27:16.0 & 26\\
OGLE-LMC-CL-1127 & LMC645.30 & SL620,LW255,KMHK1155 & C & 5:36:32.36 & -74:24:35.8 & 49\\
OGLE-LMC-CL-1128 & LMC645.31 & SL576,LW239,KMHK1092 & C & 5:33:11.50 & -74:22:29.6 & 37\\
OGLE-LMC-CL-1129 & LMC649.15 & SL737,LW312,KMHK1432 & C & 5:48:45.95 & -75:43:58.7 & 27\\
OGLE-LMC-CL-1130 & LMC649.17 & NGC2203,SL836,LW380,ESO34SC04 & C & 6:04:52.72 & -75:26:29.1 & 36\\
OGLE-LMC-CL-1131 & LMC649.23 & SL754,LW322,KMHK1473 & C & 5:50:35.80 & -75:22:16.5 & 23\\
OGLE-LMC-CL-1132 & LMC649.28 & IC2161,SL802,LW345,ESO33SC35, & C & 5:57:23.50 & -75:08:21.5 & 28\\
OGLE-LMC-CL-1133 & LMC535.10 & SL5,LW8,KMHK14 & C & 4:35:38.98 & -73:43:54.2 & 27\\
OGLE-LMC-CL-1134 & LMC543.26 & SL2,LW2,KMHK2 & C & 4:24:09.30 & -72:34:24.5 & 30\\
OGLE-LMC-CL-1135 & LMC544.08 & NGC1629,SL3,LW3,ESO55SC24,KMHK4 & C & 4:29:33.33 & -71:50:20.1 & 39\\
OGLE-LMC-CL-1136 & LMC536.08 & LW27,KMHK39 & C & 4:41:32.99 & -72:23:22.2 & 42\\
OGLE-LMC-CL-1137 & LMC536.09 & LW22,KMHK33 & C & 4:40:21.33 & -72:36:43.0 & 34\\
OGLE-LMC-CL-1138 & LMC536.10 & H88-1,KMHK16 & C & 4:36:37.89 & -72:32:11.1 & 37\\
OGLE-LMC-CL-1139 & LMC536.21 & SL4,LW4,KMHK7 & C & 4:32:40.12 & -72:20:32.5 & 42\\
OGLE-LMC-CL-1140 & LMC536.27 & KMHK19 & C & 4:37:07.05 & -72:01:15.6 & 43\\
OGLE-LMC-CL-1141 & LMC536.31 & NGC1629,SL3,LW3,ESO55SC24,KMHK4 & C & 4:29:34.26 & -71:50:13.5 & 40\\
OGLE-LMC-CL-1142 & LMC547.09 & KMHK9 & C & 4:34:55.84 & -68:14:48.7 & 28\\
OGLE-LMC-CL-1143 & LMC547.12 & KMHK3 & C & 4:29:35.23 & -68:21:22.6 & 29\\
OGLE-LMC-CL-1144 & LMC541.01 & SL42,KMHK102 & C & 4:48:11.93 & -66:46:18.7 & 27\\
OGLE-LMC-CL-1145 & LMC541.02 & KMHK93 & CA & 4:47:25.14 & -66:36:04.2 & 39\\
OGLE-LMC-CL-1146 & LMC541.03 & LW57,KMHK76 & AC & 4:46:24.02 & -66:41:33.3 & 36\\
OGLE-LMC-CL-1147 & LMC541.08 & KMHK138 & C & 4:50:43.43 & -66:13:43.5 & 54\\
OGLE-LMC-CL-1148 & LMC541.17 & BSDL81 & CA & 4:50:47.06 & -66:00:53.6 & 51\\
OGLE-LMC-CL-1149 & LMC541.25 & NGC1644,SL9,LW11,ESO84SC30, & C & 4:37:41.64 & -66:11:43.9 & 38\\
OGLE-LMC-CL-1150 & LMC548.13 & HS8,KMHK5 & C & 4:30:38.13 & -66:57:19.6 & 23\\
OGLE-LMC-CL-1151 & LMC549.02 & NGC1644,SL9,LW11,ESO84SC30, & C & 4:37:42.31 & -66:12:03.6 & 40\\
OGLE-LMC-CL-1152 & LMC589.08 & KMHK34 & C & 4:41:50.87 & -64:32:48.8 & 20\\
OGLE-LMC-CL-1153 & LMC542.09 & BSDL47 & AC & 4:49:30.33 & -65:06:07.3 & 38\\
OGLE-LMC-CL-1154 & LMC542.20 & ESO85SC03,KMHK80 & C & 4:46:55.90 & -64:50:14.3 & 27\\
OGLE-LMC-CL-1155 & LMC542.23 & SL16,KMHK41 & C & 4:42:59.89 & -64:56:57.9 & 19\\
OGLE-LMC-CL-1156 & LMC542.23 & KMHK44 & C & 4:43:27.03 & -64:53:03.3 & 20\\
OGLE-LMC-CL-1157 & LMC542.28 & KMHK92 & C & 4:47:33.77 & -64:38:44.6 & 31\\
OGLE-LMC-CL-1158 & LMC542.32 & KMHK34 & C & 4:41:50.66 & -64:32:50.4 & 17\\
OGLE-LMC-CL-1159 & LMC565.21 & LW469,KMHK1742 & C & 6:21:34.01 & -72:47:26.8 & 22\\
OGLE-LMC-CL-1160 & LMC565.30 & SL882,KMHK1728/KMHK1729 & C/AC & 6:19:04.66 & -72:23:07.0 & 24\\
OGLE-LMC-CL-1161 & LMC565.32 & SL870,LW440,KMHK1705 & C & 6:14:31.82 & -72:36:41.1 & 38\\
OGLE-LMC-CL-1162 & LMC565.32 & KMHK1702 & C & 6:13:56.38 & -72:30:19.3 & 36\\
OGLE-LMC-CL-1163 & LMC654.16 & SL835,LW379,KMHK1640 & C & 6:04:48.81 & -75:06:09.5 & 21\\
OGLE-LMC-CL-1164 & LMC572.07 & NGC2203,SL836,LW380,ESO34SC04, & C & 6:04:50.39 & -75:26:21.2 & 44\\
OGLE-LMC-CL-1165 & LMC572.15 & SL835,LW379,KMHK1640 & C & 6:04:48.91 & -75:06:08.9 & 23\\
OGLE-LMC-CL-1166 & LMC572.24 & LW373,KMHK1625 & C & 6:03:14.15 & -74:44:29.8 & 23\\
OGLE-LMC-CL-1167 & LMC572.25 & NGC2190,SL819,LW357,ESO33SC36, & C & 6:01:04.58 & -74:43:33.5 & 42\\
OGLE-LMC-CL-1168 & LMC597.02 & LW276,KMHK1269 & C & 5:43:12.50 & -63:37:02.4 & 31\\
OGLE-LMC-CL-1169 & LMC597.02 & BSDL2860 & CA & 5:43:07.20 & -63:43:36.0 & 36\\
OGLE-LMC-CL-1170 & LMC597.03 & SL649,LW269,KMHK1214 & C & 5:41:10.35 & -63:46:09.5 & 37\\
OGLE-LMC-CL-1171 & LMC597.09 & SL677,LW280,KMHK1286 & C & 5:43:55.68 & -63:22:24.6 & 26\\
OGLE-LMC-CL-1172 & LMC597.10 & KMHK1278 & C & 5:43:28.42 & -63:24:51.4 & 23\\
OGLE-LMC-CL-1173 & LMC597.20 & BSDL2771 & CA & 5:41:28.05 & -63:07:47.5 & 43\\
OGLE-LMC-CL-1174 & LMC597.21 & DES001SC19 & - & 5:40:35.41 & -63:05:59.2 & 47\\
OGLE-LMC-CL-1175 & LMC597.21 & BSDL2724 & AC & 5:40:41.05 & -62:54:32.1 & 39\\
OGLE-LMC-CL-1176 & LMC597.23 & SL604,LW251,KMHK1127 & C & 5:36:46.18 & -62:52:59.1 & 28\\
OGLE-LMC-CL-1177 & LMC597.23 & BSDL2602 & CA & 5:37:55.99 & -63:07:27.9 & 53\\
OGLE-LMC-CL-1178 & LMC597.26 & NGC2097,SL682,LW282,ESO56SC28, & C & 5:44:15.28 & -62:46:59.4 & 42\\
OGLE-LMC-CL-1179 & LMC597.27 & SL670,LW277,ESO86SC25,KMHK1268 & C & 5:43:10.12 & -62:48:42.6 & 31\\
OGLE-LMC-CL-1180 & LMC597.29 & BSDL2679 & AC & 5:39:50.42 & -62:46:10.2 & 40\\
OGLE-LMC-CL-1181 & LMC597.30 & OHSC23 & AC & 5:39:12.05 & -62:35:09.5 & 46\\
OGLE-LMC-CL-1182 & LMC597.31 & BSDL2530/BSDL2533 & CA/CA & 5:36:48.89 & -62:43:10.6 & 39\\
OGLE-LMC-CL-1183 & LMC598.01 & LW278,KMHK1265 & C & 5:43:12.53 & -62:28:31.2 & 28\\
OGLE-LMC-CL-1184 & LMC598.01 & LW277/(DES001SC21) & AC/(C) & 5:43:04.27 & -62:31:47.9 & 37\\
OGLE-LMC-CL-1185 & LMC598.02 & LW272,KMHK1241 & C & 5:42:25.34 & -62:30:08.0 & 25\\
OGLE-LMC-CL-1186 & LMC598.02 & DES001SC22 & - & 5:41:53.57 & -62:21:45.1 & 25\\
OGLE-LMC-CL-1187 & LMC598.04 & DES001SC17 & - & 5:38:58.79 & -62:26:26.6 & 30\\
OGLE-LMC-CL-1188 & LMC598.07 & BSDL2357/(DES000SC16) & A/(C) & 5:34:48.99 & -62:25:04.7 & 36\\
OGLE-LMC-CL-1189 & LMC598.23 & E2,ESO120SC08,KMHK1119 & C & 5:36:21.86 & -61:47:20.6 & 28\\
OGLE-LMC-CL-1190 & LMC593.01 & DES001SC13 & - & 5:33:37.01 & -63:04:17.4 & 46\\
OGLE-LMC-CL-1191 & LMC593.04 & OHSC18/(DES001SC06) & A/(C) & 5:29:29.04 & -63:05:00.1 & 37\\
OGLE-LMC-CL-1192 & LMC593.07 & SL448,LW205,ESO85SC82,KMHK859 & C & 5:24:59.78 & -63:02:57.4 & 28\\
OGLE-LMC-CL-1193 & LMC593.07 & LW209,KMHK877 & C & 5:25:45.10 & -63:01:17.9 & 39\\
OGLE-LMC-CL-1194 & LMC593.10 & BSDL2214 & CA & 5:32:43.88 & -62:38:33.5 & 37\\
OGLE-LMC-CL-1195 & LMC593.18 & BSDL2260 & C & 5:33:26.69 & -62:29:41.9 & 20\\
OGLE-LMC-CL-1196 & LMC575.14 & ESO119SC50,KMHK705 & C & 5:16:41.79 & -62:01:23.7 & 21\\
OGLE-LMC-CL-1197 & LMC575.18 & LW195,ESO119SC61,KMHK821 & C & 5:22:35.18 & -61:52:44.4 & 23\\
OGLE-LMC-CL-1198 & LMC574.03 & SL388,LW186,ESO85SC72,KMHK773 & C & 5:20:07.00 & -63:29:03.9 & 41\\
OGLE-LMC-CL-1199 & LMC574.08 & KMHK854 & CA & 5:24:40.70 & -63:12:15.7 & 37\\
OGLE-LMC-CL-1200 & LMC574.13 & SL354,LW177,ESO85SC63,KMHK712 & C & 5:17:34.60 & -63:25:16.5 & 50\\
OGLE-LMC-CL-1201 & LMC574.14 & SL346,LW176,ESO85SC60,KMHK702 & C & 5:16:28.26 & -63:09:58.9 & 34\\
OGLE-LMC-CL-1202 & LMC574.14 & SL345,LW175,ESO85SC58,KMHK699 & C & 5:16:18.49 & -63:12:15.4 & 39\\
OGLE-LMC-CL-1203 & LMC574.17 & SL448,LW205,ESO85SC82,KMHK859 & C & 5:24:59.95 & -63:02:56.0 & 28\\
OGLE-LMC-CL-1204 & LMC574.20 & OHSC10,KMHK782 & C & 5:20:38.47 & -63:07:52.1 & 31\\
OGLE-LMC-CL-1205 & LMC574.21 & NGC1900,SL376,LW184,ESO85SC68, & C & 5:19:08.52 & -63:01:26.2 & 39\\
OGLE-LMC-CL-1206 & LMC579.20 & SL262,LW146,ESO119SC40,KMHK582 & C & 5:09:22.54 & -62:22:43.9 & 25\\
OGLE-LMC-CL-1207 & LMC578.01 & SL303,LW158,KMHK644 & C & 5:12:21.58 & -64:18:18.9 & 41\\
OGLE-LMC-CL-1208 & LMC578.01 & H88-207,H80F4-1/H88-208,H80F4-2 & AC/A & 5:12:32.72 & -64:13:06.9 & 46\\
OGLE-LMC-CL-1209 & LMC578.03 & LW148,KMHK590 & C & 5:09:40.65 & -64:10:14.2 & 31\\
OGLE-LMC-CL-1210 & LMC578.04 & SL259,LW145,KMHK577 & C & 5:08:50.47 & -64:06:11.4 & 21\\
OGLE-LMC-CL-1211 & LMC578.08 & H88-225,H80F4-3 & AC & 5:14:29.10 & -63:57:01.8 & 47\\
OGLE-LMC-CL-1212 & LMC578.09 & SL305,LW159 & AC & 5:12:24.70 & -64:01:41.8 & 45\\
OGLE-LMC-CL-1213 & LMC578.15 & DES001SC01 & - & 5:03:16.80 & -64:01:45.6 & 28\\
OGLE-LMC-CL-1214 & LMC578.20 & SL273,KMHK597 & C & 5:10:02.76 & -63:38:21.1 & 26\\
OGLE-LMC-CL-1215 & LMC578.22 & SL233,ESO85SC45,KMHK543 & C & 5:07:04.07 & -63:38:47.1 & 27\\
OGLE-LMC-CL-1216 & LMC578.31 & SL214,LW130,ESO85SC41,KMHK512 & CA & 5:05:23.96 & -63:17:07.9 & 34\\
OGLE-LMC-CL-1217 & LMC582.06 & KMHK243 & C & 4:54:50.35 & -64:55:09.1 & 24\\
OGLE-LMC-CL-1218 & LMC582.18 & LW115,KMHK427/BSDL416 & C/AC & 5:01:10.51 & -64:15:12.3 & 43\\
OGLE-LMC-CL-1219 & LMC582.19 & BSDL387 & AC & 5:00:16.96 & -64:12:43.5 & 34\\
OGLE-LMC-CL-1220 & LMC584.05 & SL126,ESO85SC21,KMHK322 & C & 4:57:22.59 & -62:32:12.3 & 19\\
OGLE-LMC-CL-1221 & LMC662.09 & OHSC37,KMHK1762 & C & 7:07:40.47 & -69:59:01.8 & 18\\
OGLE-LMC-CL-1222 & LMC601.14 & OHSC26 & CA & 5:47:24.77 & -62:37:14.5 & 34\\
OGLE-LMC-CL-1223 & LMC603.05 & BSDL3198 & AC & 6:00:24.06 & -64:53:11.7 & 37\\
OGLE-LMC-CL-1224 & LMC564.01 & KMHK1546 & CA & 5:57:19.99 & -65:31:57.8 & 44\\
OGLE-LMC-CL-1225 & LMC564.01 & BSDL3189 & C & 5:57:26.94 & -65:20:27.8 & 40\\
OGLE-LMC-CL-1226 & LMC564.01 & BSDL3187 & AC & 5:56:40.73 & -65:19:35.0 & 61\\
OGLE-LMC-CL-1227 & LMC564.01 & KMHK1539 & C & 5:56:42.37 & -65:21:52.8 & 48\\
OGLE-LMC-CL-1228 & LMC564.02 & KMHK1527 & AC & 5:55:39.24 & -65:36:56.7 & 50\\
OGLE-LMC-CL-1229 & LMC564.03 & BSDL3165 & CA & 5:53:47.03 & -65:19:46.6 & 40\\
OGLE-LMC-CL-1230 & LMC564.04 & NGC2123,SL755,LW324,ESO86SC36 & C & 5:51:42.97 & -65:20:20.7 & 50\\
OGLE-LMC-CL-1231 & LMC564.05 & KMHK1527 & AC & 5:50:47.22 & -65:37:10.0 & 36\\
OGLE-LMC-CL-1232 & LMC564.06 & SL739,LW313,KMHK1418 & C & 5:49:59.82 & -65:30:43.2 & 41\\
OGLE-LMC-CL-1233 & LMC564.08 & LW340,KMHK1548 & C & 5:57:34.99 & -65:00:02.4 & 36\\
OGLE-LMC-CL-1234 & LMC564.10 & BSDL3181 & C & 5:55:53.05 & -65:17:01.6 & 37\\
OGLE-LMC-CL-1235 & LMC564.11 & BSDL3169/(DES001SC26) & AC/(C) & 5:54:14.05 & -65:08:54.5 & 44\\
OGLE-LMC-CL-1236 & LMC564.13 & KMHK1423 & C & 5:50:22.42 & -65:01:08.6 & 45\\
OGLE-LMC-CL-1237 & LMC564.13 & LW318,KMHK1433 & C & 5:50:42.58 & -65:18:19.0 & 26\\
OGLE-LMC-CL-1238 & LMC564.14 & SL727,LW307,KMHK1393 & C & 5:49:03.09 & -65:00:29.8 & 40\\
OGLE-LMC-CL-1239 & LMC564.14 & LW309,KMHK1396 & C & 5:49:05.04 & -65:12:13.9 & 41\\
OGLE-LMC-CL-1240 & LMC564.14 & KMHK1409 & CA & 5:49:33.36 & -65:17:07.5 & 49\\
OGLE-LMC-CL-1241 & LMC564.14 & LW304/(DES001SC23)& AC/(C) & 5:48:45.41 & -65:09:17.4 & 45\\
OGLE-LMC-CL-1242 & LMC564.15 & SL720,LW299,KMHK1373 & C & 5:47:47.41 & -65:00:37.8 & 36\\
OGLE-LMC-CL-1243 & LMC564.16 & BSDL2962 & AC & 5:46:02.51 & -65:10:48.6 & 46\\
OGLE-LMC-CL-1244 & LMC564.16 & KMHK1355 & C & 5:46:48.55 & -65:17:07.1 & 45\\
OGLE-LMC-CL-1245 & LMC564.18 & BSDL3182/(DES001SC24) & AC/(C) & 5:56:19.31 & -64:43:08.2 & 48\\
OGLE-LMC-CL-1246 & LMC564.19 & LW329,KMHK1509/(DES001SC25)& AC/(C) & 5:54:37.66 & -64:53:09.2 & 47\\
OGLE-LMC-CL-1247 & LMC564.23 & SL726,LW306,KMHK1390 & C & 5:48:59.17 & -64:43:57.9 & 27\\
OGLE-LMC-CL-1248 & LMC564.23 & BSDL3044 & AC & 5:48:43.18 & -64:43:35.1 & 32\\
OGLE-LMC-CL-1249 & LMC564.27 & LW332,KMHK1518 & C & 5:55:07.93 & -64:39:13.8 & 41\\
OGLE-LMC-CL-1250 & LMC564.31 & SL724,LW305,KMHK1388 & C & 5:48:53.38 & -64:38:02.1 & 29\\
OGLE-LMC-CL-1251 & LMC596.01 & KMHK1322 & CA & 5:45:10.06 & -64:47:52.6 & 44\\
OGLE-LMC-CL-1252 & LMC596.01 & SL696,LW286,KMHK1324 & C & 5:45:14.98 & -64:46:18.3 & 47\\
OGLE-LMC-CL-1253 & LMC596.01 & SL689,LW284,KMHK1310 & C & 5:44:44.33 & -64:59:37.1 & 49\\
OGLE-LMC-CL-1254 & LMC596.05 & HS380,KMHK1192 & CA & 5:40:02.52 & -65:00:03.1 & 42\\
OGLE-LMC-CL-1255 & LMC596.05 & BSDL2692 & A & 5:39:55.78 & -64:49:56.9 & 52\\
OGLE-LMC-CL-1256 & LMC596.06 & BSDL2596 & AC & 5:37:33.76 & -64:52:07.3 & 47\\
OGLE-LMC-CL-1257 & LMC596.07 & BSDL2562 & C & 5:36:58.63 & -64:54:39.8 & 49\\
OGLE-LMC-CL-1258 & LMC596.12 & BSDL2775 & AC & 5:41:25.88 & -64:35:59.2 & 50\\
OGLE-LMC-CL-1259 & LMC596.12 & BSDL2750 & AC & 5:40:50.75 & -64:40:14.5 & 51\\
OGLE-LMC-CL-1260 & LMC596.14 & KMHK1149 & CA & 5:37:45.02 & -64:38:39.5 & 51\\
OGLE-LMC-CL-1261 & LMC596.14 & BSDL2627/(DES001SC15)& A/(C) & 5:38:18.42 & -64:30:27.2 & 50\\
OGLE-LMC-CL-1262 & LMC596.15 & LW250,KMHK1129 & C & 5:36:38.42 & -64:23:14.0 & 33\\
OGLE-LMC-CL-1263 & LMC596.16 & BSDL2430 & A & 5:35:13.53 & -64:39:02.8 & 54\\
OGLE-LMC-CL-1264 & LMC596.21 & KMHK1195 & C & 5:40:23.85 & -64:15:00.7 & 38\\
OGLE-LMC-CL-1265 & LMC596.21 & LW266,KMHK1198 & C & 5:40:31.64 & -64:18:01.1 & 47\\
OGLE-LMC-CL-1266 & LMC596.22 & LW260,KMHK1168 & CA & 5:38:54.30 & -64:17:43.1 & 39\\
OGLE-LMC-CL-1267 & LMC596.22 & LW261,KMHK1171 & CA & 5:39:15.28 & -64:08:23.7 & 58\\
OGLE-LMC-CL-1268 & LMC596.24 & LW250,KMHK1129 & C & 5:36:38.60 & -64:23:11.3 & 31\\
OGLE-LMC-CL-1269 & LMC596.24 & BSDL2569 & AC & 5:37:21.50 & -64:07:16.2 & 39\\
OGLE-LMC-CL-1270 & LMC596.27 & BSDL2896 & C & 5:44:04.82 & -63:55:28.6 & 50\\
OGLE-LMC-CL-1271 & LMC596.27 & SL680,LW281,KMHK1290 & C & 5:44:04.78 & -63:55:27.6 & 43\\
OGLE-LMC-CL-1272 & LMC596.28 & BSDL2823 & AC & 5:42:19.15 & -63:51:52.4 & 47\\
OGLE-LMC-CL-1273 & LMC596.29 & SL649,LW269,KMHK1214 & C & 5:41:09.13 & -63:46:24.0 & 32\\
OGLE-LMC-CL-1274 & LMC592.01 & BSDL2296 & AC & 5:33:54.71 & -64:13:00.3 & 49\\
OGLE-LMC-CL-1275 & LMC592.01 & BSDL2313 & AC & 5:34:13.75 & -64:14:31.1 & 55\\
OGLE-LMC-CL-1276 & LMC592.03 & SL549,KMHK1013 & C & 5:32:00.41 & -64:14:38.3 & 38\\
OGLE-LMC-CL-1277 & LMC592.03 & BSDL2016 & A & 5:30:53.70 & -64:07:28.5 & 47\\
OGLE-LMC-CL-1278 & LMC592.05 & HS317,KMHK947 & C & 5:29:01.95 & -64:17:14.4 & 37\\
OGLE-LMC-CL-1279 & LMC592.05 & HS305, KMHK921/(DES001SC08)& AC/(C) & 5:27:54.12 & -64:10:30.1 & 40\\
OGLE-LMC-CL-1280 & LMC592.06 & BSDL1735 & CA & 5:27:15.42 & -64:15:42.2 & 43\\
OGLE-LMC-CL-1281 & LMC592.06 & BSDL1727 & A & 5:27:05.23 & -64:19:16.6 & 58\\
OGLE-LMC-CL-1282 & LMC592.07 & BSDL1586 & A & 5:25:29.32 & -64:21:34.4 & 46\\
OGLE-LMC-CL-1283 & LMC592.11 & SL540,LW232,KMHK1003 & C & 5:31:34.27 & -63:53:46.2 & 27\\
OGLE-LMC-CL-1284 & LMC592.12 & SL525,LW225,KMHK973 & C & 5:30:23.05 & -64:01:05.4 & 28\\
OGLE-LMC-CL-1285 & LMC592.13 & OHSC17,KMHK946 & C & 5:29:05.38 & -63:46:15.7 & 29\\
OGLE-LMC-CL-1286 & LMC592.16 & NGC1942,SL445,LW203,ESO85SC81, & C & 5:24:42.33 & -63:56:40.7 & 43\\
OGLE-LMC-CL-1287 & LMC592.17 & OHSC22,KMHK1089 & C & 5:34:51.23 & -63:37:54.4 & 28\\
OGLE-LMC-CL-1288 & LMC592.19 & LW238,KMHK1044 & C & 5:32:54.65 & -63:38:24.2 & 33\\
OGLE-LMC-CL-1289 & LMC592.20 & SL529,LW229,ESO86SC5,KMHK992 & C & 5:31:07.41 & -63:32:18.5 & 36\\
OGLE-LMC-CL-1290 & LMC592.20 & OHSC20 & CA & 5:31:20.01 & -63:40:15.7 & 29\\
OGLE-LMC-CL-1291 & LMC592.21 & SL509,LW221,ESO85SC91,KMHK957 & C & 5:29:48.24 & -63:39:05.1 & 44\\
OGLE-LMC-CL-1292 & LMC592.22 & OHSC17,KMHK946 & C & 5:29:04.58 & -63:46:08.7 & 34\\
OGLE-LMC-CL-1293 & LMC592.28 & LW230 & C & 5:31:23.45 & -63:15:40.9 & 35\\
OGLE-LMC-CL-1294 & LMC592.29 & SL515,LW223,ESO85SC92,KMHK965 & C & 5:30:09.08 & -63:25:35.9 & 46\\
OGLE-LMC-CL-1295 & LMC592.29 & NGC1997,SL520,LW226,ESO86SC1, & C & 5:30:31.16 & -63:11:57.6 & 34\\
OGLE-LMC-CL-1296 & LMC573.01 & BSDL1421 & AC & 5:22:57.05 & -64:56:18.9 & 51\\
OGLE-LMC-CL-1297 & LMC573.03 & HS237,KMHK785/BSDL1272 & C/AC & 5:20:30.19 & -64:48:08.6 & 47\\
OGLE-LMC-CL-1298 & LMC573.03 & BSDL1230 & A & 5:19:50.03 & -64:46:05.1 & 50\\
OGLE-LMC-CL-1299 & LMC573.03 & KMHK772 & C & 5:19:56.12 & -65:00:15.8 & 42\\
OGLE-LMC-CL-1300 & LMC573.04 & BSDL1199/BSDL1201 & CA/A & 5:19:19.87 & -64:49:27.0 & 45\\
OGLE-LMC-CL-1301 & LMC573.07 & SL329,LW166,KMHK668 & C & 5:14:12.71 & -64:48:44.6 & 38\\
OGLE-LMC-CL-1302 & LMC573.07 & HS185,KMHK677 & CA & 5:14:47.18 & -64:59:27.5 & 43\\
OGLE-LMC-CL-1303 & LMC573.07 & BSDL978 & A & 5:14:59.07 & -64:52:52.4 & 45\\
OGLE-LMC-CL-1304 & LMC573.07 & LW165,H88-228,H80F3-22,KMHK673 & CA & 5:14:26.16 & -65:00:11.4 & 53\\
OGLE-LMC-CL-1305 & LMC573.08 & BSDL1565 & A & 5:25:12.33 & -64:29:24.6 & 47\\
OGLE-LMC-CL-1306 & LMC573.09 & HS258,KMHK837 & C & 5:23:10.33 & -64:40:46.9 & 37\\
OGLE-LMC-CL-1307 & LMC573.09 & BSDL1488 & A & 5:23:48.25 & -64:36:42.2 & 45\\
OGLE-LMC-CL-1308 & LMC573.09 & BSDL1450 & A & 5:23:14.83 & -64:27:21.4 & 45\\
OGLE-LMC-CL-1309 & LMC573.09 & BSDL1404 & A & 5:22:48.60 & -64:33:04.0 & 54\\
OGLE-LMC-CL-1310 & LMC573.09 & BSDL1398 & AC & 5:22:46.84 & -64:38:18.4 & 43\\
OGLE-LMC-CL-1311 & LMC573.10 & KMHK815 & C & 5:21:58.55 & -64:34:57.7 & 44\\
OGLE-LMC-CL-1312 & LMC573.11 & HS240,KMHK797 & AC & 5:21:01.35 & -64:30:01.7 & 48\\
OGLE-LMC-CL-1313 & LMC573.12 & BSDL1207 & A & 5:19:26.43 & -64:32:34.2 & 50\\
OGLE-LMC-CL-1314 & LMC573.17 & DES001SC04/BSDL1538 & AC & 5:24:31.95 & -64:19:33.0 & 40\\
OGLE-LMC-CL-1315 & LMC573.18 & BSDL1457 & A & 5:23:15.84 & -64:20:09.0 & 47\\
OGLE-LMC-CL-1316 & LMC573.21 & SL372,LW180,KMHK730 & C & 5:18:36.43 & -64:06:45.1 & 36\\
OGLE-LMC-CL-1317 & LMC573.28 & SL401,LW190,KMHK791 & C & 5:20:50.43 & -64:00:11.0 & 29\\
OGLE-LMC-CL-1318 & LMC573.30 & H88-258,H80F4-7,KMHK720 & C & 5:17:51.46 & -63:47:45.4 & 33\\
OGLE-LMC-CL-1319 & LMC573.30 & H88-257,H80F4-6 & C & 5:17:49.09 & -64:02:44.1 & 37\\
OGLE-LMC-CL-1320 & LMC573.32 & NGC1868,SL330,LW169,ESO85SC56, & C & 5:14:37.68 & -63:57:41.0 & 52\\
OGLE-LMC-CL-1321 & LMC573.32 & H88-225,H80F4-3 & AC & 5:14:23.78 & -64:01:22.5 & 43\\
OGLE-LMC-CL-1322 & LMC513.01 & HS173,KMHK654 & C & 5:12:51.88 & -65:21:57.6 & 42\\
OGLE-LMC-CL-1323 & LMC513.01 & H88-195,H80F3-10,KMHK634 & CA & 5:11:29.30 & -65:27:00.1 & 42\\
OGLE-LMC-CL-1324 & LMC513.02 & H88-183,H80F3-7,KMHK614 & C & 5:10:41.72 & -65:35:33.0 & 50\\
OGLE-LMC-CL-1325 & LMC513.02 & HS135,KMHK599 & C & 5:10:00.84 & -65:25:31.9 & 35\\
OGLE-LMC-CL-1326 & LMC513.03 & BSDL731 & A & 5:09:35.56 & -65:19:54.6 & 50\\
OGLE-LMC-CL-1327 & LMC513.04 & H88-141,H80F2-12 & C & 5:07:33.27 & -65:19:58.2 & 49\\
OGLE-LMC-CL-1328 & LMC513.05 & LW135,KMHK539 & C & 5:06:42.17 & -65:37:23.5 & 43\\
OGLE-LMC-CL-1329 & LMC513.06 & KMHK519 & CA & 5:05:22.27 & -65:28:59.6 & 34\\
OGLE-LMC-CL-1330 & LMC513.07 & SL193,LW122,KMHK474 & C & 5:03:27.82 & -65:22:37.9 & 43\\
OGLE-LMC-CL-1331 & LMC513.07 & SL195,LW123,KMHK476 & C & 5:03:31.58 & -65:26:19.0 & 46\\
OGLE-LMC-CL-1332 & LMC513.07 & LW124,KMHK481 & C & 5:03:38.09 & -65:34:01.4 & 48\\
OGLE-LMC-CL-1333 & LMC513.08 & H88-215,H80F3-15 & C & 5:12:58.43 & -65:05:50.0 & 47\\
OGLE-LMC-CL-1334 & LMC513.09 & NGC1859,SL297,LW156,ESO85SC50, & C & 5:11:32.94 & -65:14:43.6 & 45\\
OGLE-LMC-CL-1335 & LMC513.09 & H88-200,H80F3-11 & AC & 5:11:45.89 & -65:07:13.0 & 50\\
OGLE-LMC-CL-1336 & LMC513.09 & H88-201,H80F3-12 & AC & 5:11:47.15 & -65:05:47.3 & 52\\
OGLE-LMC-CL-1337 & LMC513.09 & LW157,KMHK637 & CA & 5:11:53.78 & -65:00:10.9 & 50\\
OGLE-LMC-CL-1338 & LMC513.10 & H88-176,H80F3-5,KMHK607/BSDL763 & C/A & 5:10:18.42 & -65:15:54.0 & 46\\
OGLE-LMC-CL-1339 & LMC513.10 & BSDL769 & AC & 5:10:06.06 & -65:15:41.6 & 22\\
OGLE-LMC-CL-1340 & LMC513.10 & KMHK620/BSDL787/BSDL785 & AC/A/A & 5:10:56.63 & -65:00:02.0 & 42\\
OGLE-LMC-CL-1341 & LMC513.11 & H88-168,H80F3-2,KMHK595 & C & 5:09:43.44 & -65:18:17.3 & 37\\
OGLE-LMC-CL-1342 & LMC513.11 & H88-160,H80F2-20 & CA & 5:09:03.43 & -65:12:36.2 & 54\\
OGLE-LMC-CL-1343 & LMC513.11 & BSDL742 & A & 5:09:50.06 & -65:14:52.7 & 48\\
OGLE-LMC-CL-1344 & LMC513.12 & H88-138,H80F2-11/BSDL643 & A/A & 5:07:10.14 & -65:17:01.4 & 55\\
OGLE-LMC-CL-1345 & LMC513.12 & H88-143,H80F2-14/BSDL659 & CA/A & 5:07:47.53 & -65:06:48.4 & 52\\
OGLE-LMC-CL-1346 & LMC513.13 & LW134,KMHK535 & C & 5:06:27.36 & -65:11:12.7 & 48\\
OGLE-LMC-CL-1347 & LMC513.13 & H88-134,H80F2-9 & C & 5:06:53.38 & -65:15:41.4 & 44\\
OGLE-LMC-CL-1348 & LMC513.13 & H88-124,H80F2-5 & A & 5:05:52.32 & -65:16:02.7 & 50\\
OGLE-LMC-CL-1349 & LMC513.14 & LW127,KMHK504 & C & 5:04:52.06 & -65:07:36.6 & 40\\
OGLE-LMC-CL-1350 & LMC513.17 & SL329,LW166,KMHK668 & C & 5:14:11.95 & -64:48:47.5 & 32\\
OGLE-LMC-CL-1351 & LMC513.17 & HS176 & CA & 5:13:43.28 & -64:54:11.0 & 49\\
OGLE-LMC-CL-1352 & LMC513.18 & HS170 & C & 5:12:43.06 & -64:51:10.0 & 43\\
OGLE-LMC-CL-1353 & LMC513.18 & LW157,KMHK637 & CA & 5:11:52.82 & -64:59:22.3 & 47\\
OGLE-LMC-CL-1354 & LMC513.19 & SL275,LW150,KMHK605 & C & 5:10:15.86 & -64:55:21.2 & 47\\
OGLE-LMC-CL-1355 & LMC513.19 & HS143,KMHK602 & AC & 5:10:12.82 & -64:58:11.2 & 48\\
OGLE-LMC-CL-1356 & LMC513.19 & KMHK620/BSDL787 & AC/A & 5:10:57.03 & -64:59:58.3 & 45\\
OGLE-LMC-CL-1357 & LMC513.20 & SL258,LW142,KMHK571 & C & 5:08:35.87 & -64:54:29.7 & 34\\
OGLE-LMC-CL-1358 & LMC513.20 & H88-156,H80F2-18 & C & 5:08:56.94 & -64:46:26.1 & 50\\
OGLE-LMC-CL-1359 & LMC513.20 & HS125,KMHK563 & C & 5:08:27.29 & -64:52:19.0 & 59\\
OGLE-LMC-CL-1360 & LMC513.21 & H88-142,H80F2-13 & AC & 5:07:38.04 & -64:59:06.4 & 51\\
OGLE-LMC-CL-1361 & LMC513.21 & H88-153,H80F2-17 & C & 5:08:18.02 & -64:57:18.0 & 52\\
OGLE-LMC-CL-1362 & LMC513.22 & NGC1831,SL227,LW133,ESO85SC44, & C & 5:06:16.13 & -64:54:47.8 & 53\\
OGLE-LMC-CL-1363 & LMC513.22 & H88-123,H80F2-6 & C & 5:06:52.58 & -65:00:02.2 & 8\\
OGLE-LMC-CL-1364 & LMC513.26 & BSDL818 & CA & 5:11:42.27 & -64:39:12.4 & 41\\
OGLE-LMC-CL-1365 & LMC513.27 & SL287,LW152,KMHK615/BSDL778/(DES001SC02)& AC/A/(C) & 5:10:50.03 & -64:31:46.6 & 38\\
OGLE-LMC-CL-1366 & LMC513.29 & SL238,LW139,KMHK548 & C & 5:07:22.54 & -64:25:49.7 & 46\\
OGLE-LMC-CL-1367 & LMC513.29 & H88-147,H80F2-16 & AC & 5:08:00.17 & -64:39:26.5 & 50\\
OGLE-LMC-CL-1368 & LMC513.29 & H88-145,H80F2-15 & AC & 5:07:49.76 & -64:39:44.8 & 50\\
OGLE-LMC-CL-1369 & LMC520.01 & KMHK1109 & CA & 5:35:23.11 & -65:29:17.6 & 50\\
OGLE-LMC-CL-1370 & LMC520.01 & LW245,KMHK1104 & C & 5:35:06.76 & -65:19:53.1 & 45\\
OGLE-LMC-CL-1371 & LMC520.01 & BSDL2350 & AC & 5:34:24.24 & -65:22:40.4 & 51\\
OGLE-LMC-CL-1372 & LMC520.02 & KMHK1063 & CA & 5:33:37.45 & -65:22:05.4 & 49\\
OGLE-LMC-CL-1373 & LMC520.02 & BSDL2298 & C & 5:33:17.03 & -65:33:13.2 & 55\\
OGLE-LMC-CL-1374 & LMC520.02 & BSDL2256 & A & 5:33:00.25 & -65:25:30.1 & 51\\
OGLE-LMC-CL-1375 & LMC520.03 & KMHK1012 & C & 5:31:51.52 & -65:26:40.9 & 40\\
OGLE-LMC-CL-1376 & LMC520.03 & BSDL2124 & A & 5:31:48.77 & -65:20:50.8 & 52\\
OGLE-LMC-CL-1377 & LMC520.03 & BSDL2056 & CA & 5:31:20.28 & -65:19:41.4 & 59\\
OGLE-LMC-CL-1378 & LMC520.03 & BSDL2118 & CA & 5:31:45.25 & -65:19:42.7 & 47\\
OGLE-LMC-CL-1379 & LMC520.04 & LW228,KMHK990 & C & 5:30:53.49 & -65:35:07.2 & 35\\
OGLE-LMC-CL-1380 & LMC520.04 & BSDL1988 & A & 5:30:29.49 & -65:35:59.3 & 52\\
OGLE-LMC-CL-1381 & LMC520.05 & BSDL1903 & AC & 5:29:14.12 & -65:22:48.9 & 53\\
OGLE-LMC-CL-1382 & LMC520.06 & KMHK908 & AC & 5:27:01.69 & -65:27:03.9 & 47\\
OGLE-LMC-CL-1383 & LMC520.06 & BSDL1786 & AC & 5:27:38.02 & -65:23:02.7 & 49\\
OGLE-LMC-CL-1384 & LMC520.06 & HS297 & AC & 5:26:44.38 & -65:19:44.9 & 53\\
OGLE-LMC-CL-1385 & LMC520.07 & BSDL1658 & CA & 5:26:19.93 & -65:29:27.7 & 50\\
OGLE-LMC-CL-1386 & LMC520.08 & LW248,KMHK1126 & C & 5:36:25.13 & -65:16:14.0 & 35\\
OGLE-LMC-CL-1387 & LMC520.08 & BSDL2552 & AC & 5:36:56.94 & -65:11:10.5 & 45\\
OGLE-LMC-CL-1388 & LMC520.08 & BSDL2466 & AC & 5:35:37.59 & -65:02:29.4 & 69\\
OGLE-LMC-CL-1389 & LMC520.09 & LW246,KMHK1103 & C & 5:35:07.08 & -65:12:07.1 & 39\\
OGLE-LMC-CL-1390 & LMC520.09 & SL578,LW242,KMHK1080 & C & 5:34:19.29 & -65:17:19.5 & 38\\
OGLE-LMC-CL-1391 & LMC520.09 & LW244,KMHK1081 & C & 5:34:21.76 & -65:03:34.9 & 46\\
OGLE-LMC-CL-1392 & LMC520.09 & BSDL2387 & A & 5:34:50.63 & -65:09:52.4 & 46\\
OGLE-LMC-CL-1393 & LMC520.10 & BSDL2239 & AC & 5:32:49.24 & -65:08:54.9 & 45\\
OGLE-LMC-CL-1394 & LMC520.11 & KMHK1011/BSDL2135 & C/AC & 5:31:48.16 & -65:15:49.7 & 46\\
OGLE-LMC-CL-1395 & LMC520.11 & BSDL2142 & AC & 5:31:53.84 & -65:00:18.1 & 44\\
OGLE-LMC-CL-1396 & LMC520.12 & BSDL2042/BSDL2044/BSDL2046 & CA/A/AC & 5:30:58.13 & -65:02:19.8 & 51\\
OGLE-LMC-CL-1397 & LMC520.13 & SL496,LW218,KMHK935 & C & 5:28:32.45 & -65:10:28.6 & 35\\
OGLE-LMC-CL-1398 & LMC520.13 & BSDL1841 & A & 5:28:31.45 & -65:15:00.1 & 53\\
OGLE-LMC-CL-1399 & LMC520.14 & LW215,KMHK919 & C & 5:27:42.80 & -65:17:50.7 & 43\\
OGLE-LMC-CL-1400 & LMC520.16 & BSDL1537 & A & 5:24:36.85 & -65:09:50.3 & 50\\
OGLE-LMC-CL-1401 & LMC520.16 & LW199,KMHK848 & CA & 5:24:08.87 & -65:07:35.2 & 51\\
OGLE-LMC-CL-1402 & LMC520.19 & SL573,LW240,ESO86SC11,KMHK1066 & C & 5:33:46.95 & -64:56:28.2 & 39\\
OGLE-LMC-CL-1403 & LMC520.20 & H4,SL556,LW237,ESO86SC9,KMHK1034 & C & 5:32:26.30 & -64:44:11.3 & 47\\
OGLE-LMC-CL-1404 & LMC520.20 & BSDL2152 & CA & 5:32:00.13 & -64:50:26.5 & 48\\
OGLE-LMC-CL-1405 & LMC520.22 & SL500,LW219,KHMK937 & C & 5:28:39.49 & -64:51:13.1 & 40\\
OGLE-LMC-CL-1406 & LMC520.23 & BSDL1791 & A & 5:27:55.66 & -64:42:03.8 & 46\\
OGLE-LMC-CL-1407 & LMC520.24 & LW208,KMHK878 & C & 5:25:34.85 & -64:45:33.0 & 46\\
OGLE-LMC-CL-1408 & LMC520.25 & HS276/BSDL1553 & C/A & 5:24:59.15 & -64:55:38.5 & 43\\
OGLE-LMC-CL-1409 & LMC520.25 & BSDL1543 & AC & 5:24:48.73 & -64:52:01.1 & 52\\
OGLE-LMC-CL-1410 & LMC520.28 & BSDL2192 & AC & 5:32:26.34 & -64:38:24.9 & 39\\
OGLE-LMC-CL-1411 & LMC520.29 & SL511,LW222,KMHK959 & C & 5:29:48.82 & -64:26:17.6 & 40\\
OGLE-LMC-CL-1412 & LMC520.30 & KMHK938 & CA & 5:28:43.99 & -64:40:45.6 & 39\\
OGLE-LMC-CL-1413 & LMC520.31 & SL484,LW216,KMHK918 & C & 5:27:46.35 & -64:38:55.5 & 45\\
OGLE-LMC-CL-1414 & LMC520.31 & BSDL1780 & AC & 5:27:39.14 & -64:36:33.7 & 53\\
OGLE-LMC-CL-1415 & LMC520.32 & LW212,KMHK882/BSDL1630 & C/c & 5:26:09.45 & -64:33:34.7 & 33\\
OGLE-LMC-CL-1416 & LMC520.32 & BSDL1654 & A & 5:26:38.81 & -64:26:08.9 & 31\\
OGLE-LMC-CL-1417 & LMC596.16 & DES001SC14 & - & 5:35:12.87 & -64:37:22.2 & 53\\
OGLE-LMC-CL-1418 & LMC596.29 & DES001SC18 & - & 5:40:41.98 & -63:49:07.4 & 41\\
OGLE-LMC-CL-1419 & LMC592.13 & DES001SC07 & - & 5:28:44.41 & -64:03:24.7 & 41\\
OGLE-LMC-CL-1420 & LMC592.17 & DES001SC11 & - & 5:35:05.67 & -63:32:17.7 & 38\\
OGLE-LMC-CL-1421 & LMC592.28 & DES001SC09 & - & 5:31:49.44 & -63:19:19.7 & 39\\
OGLE-LMC-CL-1422 & LMC592.28 & DES001SC10 & - & 5:31:20.16 & -63:26:00.5 & 63\\
OGLE-LMC-CL-1423 & LMC520.23 & DES001SC05 & - & 5:27:41.37 & -64:48:26.1 & 51\\
OGLE-LMC-CL-1424 & LMC520.26 & DES001SC14 & - & 5:35:13.07 & -64:37:22.3 & 52\\
\hline
\end{longtable}
\end{center}
} 

The catalog, the list of all analyzed LMC fields and all the graphical
materials are avaliable on the OGLE web page:\\ \centerline{\it
  http://ogle.astrouw.edu.pl}

\section{Conclusions}

We have presented a catalog of star clusters in the outer regions of
the Large Magellanic Cloud based on the OGLE-IV deep photometric maps.
We found a total of 679 star clusters, including 226 new objects which
were not listed in any of the previous catalogs, 438 clusters listed in Bica
\etal 2008 and 15 objects listed in the Dark Energy
Survey publication (Pieres \etal 2016). For all of them the
equatorial coordinates and cross-identification with previous catalogs
are provided. The detection method presented in this paper is 
very effective. With our algorithm we found almost all previously known clusters in
this characteristic sparse region of the LMC and increased the total
number of these objects by 50\%. This paper is the first of a
series of publications. In the coming papers we will present a complete catalog of
star clusters in the whole Magellanic System with their 
parameters determined, for the first time, from a single, homogeneous
photometric data set collected by the OGLE project.

\section{Acknowledgments}

We gratefully acknowledge 
the financial support of the Polish National Science Center, 
grant SONATA 2013/11/D/ST9/03445 to D.\ Skowron.
Z.\ Kostrzewa-Rutkowska acknowledges support from European Research
Council Consolidator Grant 647208.
We would like to thank Profs.\ M. Kubiak and G. Pietrzy{\'n}ski, former
members of the OGLE team, for their contribution to the collection of
the OGLE photometric data over the past years.
The OGLE project has received funding from the National Science
Center, Poland, grant MAESTRO 2014/14/A/ST9/00121 to A.\ Udalski.

\vspace{1.5cm}
\begin{center}
References
\end{center}

\noindent
\begin{itemize}
\leftmargin 0pt
\itemsep -5pt
\parsep -5pt
\refitem{Bhatia, R. K., MacGillivray, H. T.}{1989}{ AA}{211}{9}
\refitem{Bica, E., Bonatto, C., Dutra, C. M., Santos, J. F. C.}{2008} {MNRAS}{ 389}{ 678}
\refitem{Choundhury, S., Subramaniam, A., Piatti, A. E.}{2015}{AJ}{149}{52}
\refitem{Cioni, M.-R. L. Habing, H. J., Isreal, F. P.}{2000}{AA}{358}{L9-L12}
\refitem{Deb, S. and Singh, H. P.}{2014}{VizieR Online Data Catalog}{743}{~}
\refitem{Glatt, K., Grebel, E. K., Koch, A.,}{2010}{AA}{535}{A115}
\refitem{Jacyszyn-Dobrzeniecka, A., \etal}{2016}{Acta Astr.}{66}{149}
\refitem{Klein, C. R., Cenko, S. B., Miller, A. A., Norman, D. J., Bloom, J. S.}{2014}{arXiv}{1405.1035}{~}
\refitem{Nayak, P. K., Subramaniam, A., Choudhury, S., Indu, G., Sagar, Ram}{2016} {MNRAS}{10.1093/mnras/stw2043}{~}
\refitem{Palma, T., Gramajo, L. V., Claria, J. J., Lares, M., Geisler, D., Ahumada, A. V.}{2016} {AA}{586}{A41}
\refitem{Piatti, A. E., Sarajedini, A., Geisler, D., Bica, E., Claria, J. J.}{2002} {MNRAS}{329}{556}
\refitem{Piatti, A. E., Geisler, D., Bica, E., Claria, J. J.}{2003b} {MNRAS}{343}{851}
\refitem{Piatti, A. E., Bica, E., Geisler, D., Claria, J. J.}{2003a} {MNRAS}{344}{965}
\refitem{Piatti, A. E., Geisler, D., Sarajedini, A., Gallart, C.}{2009} {AA}{501}{585}
\refitem{Piatti, A. E., \etal}{2014} {AA}{570}{A74}
\refitem{Piatti, A. E., \etal}{2015} {MNRAS}{454}{839}
\refitem{Pieres, A., \etal}{2016} {MNRAS}{461}{519}
\refitem{Pietrzy{\'n}ski, G. Udalski, A., Kubiak, M., Szyma{\'n}ski, M. K., Wozniak, P., Zebrun, K.}{1999}{Acta Astr.}{49}{521}
\refitem{Pietrzy{\'n}ski, G., Udalski, A.}{2000}{Acta Astr.}{50}{337}
\refitem{Schechter, P. L., Mateo, M., Saha, A.}{1993}{PASP}{105}{1342}
\refitem{Scowcroft, V., Freedman, W. L., Madore, B. F., Monson, B. F., Monson, A., Persson, S. E., Rich, J., Seibert, M., Rigby, J. R.}{2016}{ApJ}{816}{10pp}
\refitem{Skowron, D. M. \etal}{2014}{ApJ}{795}{17}
\refitem{Udalski, A., Kubiak, M., Szyma{\'n}ski, M. K.,}{1997} {Acta Astron.}{47}{319}
\refitem{Udalski, A., Szyma{\'n}ski, M. K., Soszy{\'n}ski, I., Poleski, R.}{2008} {Acta Astron.}{58}{69}
\refitem{Udalski, A., Szyma{\'n}ski, M. K., Szyma{\'n}ski, G.}{2015} {Acta Astron.}{ 65}{ 1}
\refitem{Zaritsky, D., Harris, J., Thompson, I.}{1997}{Astron.J.}{114}{1002}
\end{itemize}

\end{document}